\documentstyle{article}

\newcommand{\Dim}{{\rm dim}\;}
\newtheorem{theo}{Theorem}[section]
\newtheorem{lem}[theo]{Lemma}
\newtheorem{cor}[theo]{Corollary}
\newtheorem{con}[theo]{Conjecture}
\date{}
\author{M. BAUER and N. SOCHEN \\ {\em Service de physique th\'eorique de
Saclay$^1$} \\ {\em F-91191 Gif-sur-Yvette cedex,
France}}
\title{Fusion and singular vectors in $A_1^{(1)}$ highest weight cyclic modules
}
\begin{document}
\maketitle
\addtocounter{footnote}{1}
\footnotetext{Laboratoire de la Direction des Sciences de la Mati\`ere du
Commissariat \`a l'{\cal E}nergie Atomique}

\begin{abstract}
We show how the interplay between the fusion formalism of conformal field
theory and the Knizhnik--Zamolodchikov equation leads to explicit formul\ae\
for the singular vectors in the highest weight representations of $A_1^{(1)}$.
\end{abstract}

\section{Introduction} \label{sec:i}

Infinite dimensional Lie algebras occur everywhere in the study of 2-d
conformal field theories: the Virasoro algebra and the affine algebras are the
most common examples.
However the construction of the irreducible representations of these algebras
is quite involved.
Singular vectors are important because they indicate the existence of
subrepresentations in a given representation.
In the affine case,
Kac and Kazhdan \cite{kk}  gave the criterion for the reducibility or
irreducibility of the Verma modules and Malikov,
Feigin and Fuks \cite{mff} found a formula for the singular vectors in the
$A_N^{(1)}$ case.
This formula looks very simple,
but involves an analytic continuation to make sense,
which makes it very difficult to use.

Apart from the purely mathematical description,
several approaches motivated by physics have been proposed,
based on vertex operators (see \cite{tk} for a general reference dealing with
$A_1^{(1)}$), bosonization and variants of the Feigin and Fuks construction and
BRST cohomology \cite{bf}.
In the physical context,
the importance of singular vectors comes from Ward identities: to calculate a
correlation function involving a descendent of a primary field,
one simply applies a linear operator to the correlation function of the primary
\cite{bpz}.
A singular vector is a descendent that is set to zero in an irreducible
representation,
with the consequence that the correlation functions of the corresponding
primary satisfy closed linear relations,
leading to a contour integral representation.

One of the aims of this paper is to show that elementary methods of conformal
field theory allow us to understand some important features of the structure of
representations of theses algebras.
Our inspiration comes from remarks at the end of the seminal paper of Belavin,
Polyakov and Zamolodchikov (see appendix B in \cite{bpz}).
We restricted our attention to the $A_1^{(1)}$ algebra not only for simplicity
(although generalization is not straightforward,
we believe that the same methods applied to other affine algebras will lead to
interesting results),
but also because we hoped to get a better understanding of the construction
made in \cite{bfiz}  by Bauer,
Di Francesco,
Itzykson and Zuber for the singular vectors in Virasoro Verma modules.

The basic idea is the following : the symmetries of conformal field theories
are so large that they determine ``almost'' completely the structure of the
operator product expansion of primary fields.
A remarkable homogeneous linear system, the system of descent equations (see
section \ref{sec:fde}), encodes this structure.
The singular vectors are in the kernel of the descent equations, and by
duality, they also appear as an obstruction to solve the linear system.
This can be used to compute them.

The $A_1^{(1)}$ case has its own peculiarities,
but is in a sense easier to deal with than the case of the Virasoro algebra,
and a more complete treatment is possible.
We still expect a precise connection between the two cases via Hamiltonian
reduction \cite{bo},
although as yet we have only been able to work out some simple examples.

\vspace{5 mm}

The organization of this paper is as follows.
We begin with a short reminder of the basic notions in the representation
theory of affine algebras in our particular case. We introduce Verma modules,
singular vectors, and the contragredient form.  This is standard material,
included only for the sake of completeness. For a more detailed and pedagogical
presentation, see \cite{kr}.
The next section quotes (again restricting to the $A_1^{(1)}$ case) the results
of Kac and Kazhdan \cite{kk},
and  the formula for singular vectors given by Malikov,
Feigin and Fuks \cite{mff}.
We decided to include some of the proofs,
hoping that a physicist's style could make them accessible to a larger
audience.
Furthermore,
some features of our constructions have counterparts in these proofs,
showing clearly that for the time being,
our work is not a substitute for the usual representation theory,
but uses it in several places.
In section \ref{sec:pff} we introduce the notion of Verma primary fields and
explain fusion from a naive point of view.
This leads to the ``descent equations'', which summarize the structure of the
operator product expansion.
We end this section with some comments showing the relation with a more
mathematical definition of fusion.
In section \ref{sec:frde} we derive important consequences of the descent
equations, using the contragredient form as a fundamental tool. This leads to
the existence of fusion rules.
In section \ref{sec:srde} we recast the descent equations in triangular form,
and point out the role played by the so-called Knizhnik--Zamolodchikov
equation. This allows us to calculate recursively all the descendants of a
primary field in a fusion process.
We use this recursive form in section \ref{sec:sv} to obtain explicit recursion
relations or matrix forms to calculate the singular vectors.
The next section is devoted to some simple comments related to our initial
motivations,
i.e. the relation with the case of the Virasoro algebra via Hamiltonian
reduction.
Some technical details are treated in appendix.
We have tried to give a self-contained and pedagogical presentation,
but decided to refer systematically to \cite{bfiz} for the comparisons with the
case of the Virasoro algebra.

\section{Basic definitions} \label{sec:bd}

\subsection{The $A_1^{(1)}$ algebra}

The $A^{(1)}_1$ algebra (which we shall also denote simply by ${\cal A}$) can
be presented as a current algebra with generators $k$ and $J_n^a$,
$n \in {\bf Z}$,
$a \in \{-,0,+\}$ satisfying the following commutation relations:
\[
[J^+_m,J^+_n]=0 \qquad [J^-_m,J^-_n]=0 \qquad [J^a_m,k]=0  \qquad
[J^0_m,J^+_n]= J^+_{m+n} \qquad [J^0_m,J^-_n]=-J^-_{m+n}
\]
\begin{equation} \label{eq:com}
[J^0_m,J^0_n]=\frac{k}{2}m\delta_{n+m}   \qquad
[J^+_m,J^-_n]=km\delta_{n+m}+2J^0_{m+n}
\end{equation}
This algebra is doubly graded if we define
\[ \overline{d}(J^a_n)=a \quad \overline{d}(k)=0 \qquad \underline{d}(J^a_n)=n
\quad \underline{d}(k)=0 \]
The so-called principal gradation $d=2\underline{d} +\overline{d}$ is used to
define several subalgebras needed to construct the $A^{(1)}_1$ Verma modules.
We remark that the commutation relations with $J^0_0$ simply calculate the
$\overline{d}$ gradation i.e. $ad(J^0_0)$ is multiplication by $\overline{d}$.
It is also useful to add to ${\cal A}$ a generator called $\underline{D}$ with
analogous properties with respect to $\underline{d}$,
that is
\[ [\underline{D},J^a_n]=nJ^a_n \qquad [\underline{D},k]=0 \]
The Jacobi identities are still true because ${\cal A}$ is graded by
$\underline{d}$.
Shifting  $\underline{D}$ by a constant does not change the commutation
relations.
We set $\hat{\cal A}={\cal A} \oplus {\bf C}\underline{D}$.
In physical applications,
the Sugawara construction will provide an explicit form for $\underline{D}$ so
adding it to ${\cal A}$ is not completely artificial.
Up to an additive constant,
$-\underline{D}$ will be the energy operator,
which we require to be bounded below in representations.

We write
\[\hat{\cal A}=\bigoplus_{i \in {\bf Z}} {\cal E}_i = {\cal E}_- \oplus {\cal
E}_0 \oplus {\cal E}_+\]
where ${\cal E}_i$ is the subspace on which $d=2\underline{d} +\overline{d}$
takes the value $i$ and ${\cal E}_-$ (resp. ${\cal E}_+$) is the direct sum of
the ${\cal E}_i$'s for negative (resp. positive) $i$'s.
Finally we let ${\cal B}={\cal E}_0\oplus{\cal E}_+$.
The dimension of ${\cal E}_0$ is 3 and the dimension of ${\cal E}_i$,
$i \neq 0$ is 1 or 2 depending on whether $i$ is even or odd.
It is easy to check that the smallest Lie subalgebra of ${\cal A}$ containing
${\cal E}_{-1}$ (resp. ${\cal E}_1$) is ${\cal E}_-$  (resp. ${\cal E}_+$).
Furthermore ${\cal E}_{-1} \oplus {\cal E}_1$ generates ${\cal A}$.
This last observation can be generalized (see  \cite{k}) to give an axiomatic
definition of affine algebras by generators and relations,
leading to a theory very akin to the theory of finite dimensional complex
semi-simple Lie algebras.

\vspace{7 mm}

We introduce now the basic tools to study a certain class of representations of
$\hat{\cal A}$. We begin by recalling some useful concepts. For the rest of
this section, we more or less follow \cite{kr}.

\subsection{Verma modules} \label{sec:Vm}

Let ${\cal G}$ be a Lie algebra.
We shall denote by $U({\cal G})$ its universal enveloping algebra.
This space can be defined abstractly as the quotient of the tensor algebra over
${\cal G}$ by the two-sided ideal generated by the commutation relations in
$G$.
This definition is just what is needed to make  representations of ${\cal G}$
and left  $U({\cal G})$-modules the same thing.
Naively,
when we make calculations in a representation of ${\cal G}$ on a space $E$ we
manipulate the representatives of elements of ${\cal G}$ in $End(E)$,
and $U({\cal G})$ is the space where we can make all the manipulations which do
not really depend on the particular representation we are dealing with but only
on the commutation relations in ${\cal G}$.
We now state two results which we shall need later on.
\begin{itemize}
\item The first one is the Poincar\'e-Birkhoff-Witt theorem: fix a basis
${\gamma}_i$ of ${\cal G}$ as a vector space,
where $i$ belongs to some ordered set $I$,
then monomials of the form ${\gamma}_{i_1} \cdots {\gamma}_{i_n} $,
where $i_1 \leq \cdots \leq i_n$,
form a basis of $U({\cal G})$ as a vector space.
The hard part is of course the fact that these monomials are linearly
independent.
To see that they span $U({\cal G})$ we simply apply the commutation relations.
In the special case when the algebra we deal with  is an oscillator algebra
this simply tells us that it is possible to put the annihilation operators on
the left and the creation operators on the right by applying the commutation
relations (in fact we shall see in the next section that ${\cal A}$ is not too
different from an oscillator algebra and use an interesting consequence of this
fact).
\item The second one is the fact that $U({\cal E}_-)$ does not contain zero
divisors.
\end{itemize}
For an elementary and lucid account on universal enveloping algebras, see
\cite{kna}.

\vspace{7 mm}

Verma modules are usually defined by giving properties that characterize them.
The starting point is a one dimensional representation of ${\cal E}_0$,
a maximal Abelian subalgebra of $\hat{\cal A}$.
In this representation, $J^0_0$ and $k$ act by scalars which we denote
generically by $\jmath$ and $t-2$.
By analogy with the finite dimensional Lie algebra $A_1$, we shall sometimes
call $\jmath$ the spin of the representation.
The value of $\underline{D}$ is immaterial,
we take it to be 0.
We can turn this space into a one dimensional representation of $B$ by letting
${\cal E}_+$ act as $0$. We denote this representation of ${\cal B}$ by ${\bf
C}^{(\jmath,t)}$. A Verma module  $V^{(\jmath,t)}$ for $\hat{\cal A}$ is a
representation of $\hat{\cal A}$ with the following properties:
\begin{enumerate}
 \item The module $V^{(\jmath,t)}$ contains a one dimensional subspace
$V_{0,0}$ carrying a representation of ${\cal B}$ isomorphic to  ${\bf
C}^{(\jmath,t)}$.
\item The smallest subspace of $V^{(\jmath,t)}$ stable under the action of
$\hat{\cal A}$ and containing $V_{0,0}$ is  $V^{(\jmath,t)}$ itself.
\item Any representation of $\hat{\cal A}$ satisfying the first two properties
is isomorphic to a quotient of  $V^{(\jmath,t)}$.
\end{enumerate}
These properties make it clear that two Verma modules associated with the same
${\bf C}^{(\jmath,t)}$ are canonically isomorphic,
so Verma modules if they exist are unique.
Usually,
representations satisfying properties one and two are called cyclic
representations.

To prove existence we consider the induced representation $U(\hat{\cal A})
\otimes_{U({\cal B})} {\bf C}^{(\jmath,t)}$.
As an $U(\hat{\cal A})$-module this is isomorphic to the quotient of
$U(\hat{\cal A})$ by the left ideal generated by $J^0_0-\jmath$,
$k-(t-2)$,
$\underline{D}$,
and the $J^a_n$'s in ${\cal E}_+$.
We denote this ideal by ${\cal I}^{(\jmath,t)}$.
It is easy to check properties 1,
2 and 3 for this representation.
We can now order the generators according to the principal gradation and apply
the Poincar\'e-Birkhoff-Witt theorem to see that any element in $U(\hat{\cal
A})$
can be written as a linear combination of terms of the form $x_-x_0x_+$ with
$x_a \in U({\cal E}_a)$ for $a \in \{-,0,+\}$.
This implies that $V^{(\jmath,t)}$ is isomorphic to  $U({\cal E}_-)$ as an
$U({\cal E}_-)$-module.
If $x \in U(\hat{\cal A})$ we denote its image in the quotient by $|x \rangle$.
The module property is simply that $x|y \rangle = |xy \rangle $,
and we call $|1 \rangle$ the highest weight vector,
a terminology borrowed from the theory of semi-simple Lie algebras.
Later,
when we need to manipulate several Verma modules at the same time,
we shall use the notation $|\jmath,t\rangle$ for the highest weight vector in
$V^{(\jmath,t)}$.

\vspace{7 mm}

Let us finally remark that $V^{(\jmath,t)}$ is a doubly graded representation.
In fact the Poincar\'e-Birkhoff-Witt theorem  implies that the monomials
\begin{equation} \label{eq:pbw} \prod_{i=1}^{+\infty} (J^+_{-i})^{p_{i,+}}
\prod_{i=1}^{+\infty} (J^0_{-i})^{p_{i,0}} \prod_{i=0}^{+\infty}
(J^-_{-i})^{p_{i,-}} |1 \rangle \end{equation}
(where all but a finite number of the integers $p$'s are zero) form a basis of
the Verma module.
The values of $-\underline{d}$ and $-\overline{d}$  on such a monomial are
respectively $n=\sum_{i,a} ip_{i,a}$ and $m=-\sum_{i,a} ap_{i,a}$,
and we see that $n$ is always non-negative and $m$ is never less than $-n$.
We denote by $I$ (see figure~1) the set of couples $(n,m)$ and end up with a
decomposition
\[ V^{(\jmath,t)}=\bigoplus_{(n,m) \in I} V_{n,m} \]
\begin{figure}
\begin{picture}(400,200)(-55,0)
\multiput(20,20)(40,0){10}{\circle*{7}}
\multiput(60,60)(40,0){9}{\circle*{7}}
\multiput(100,100)(40,0){8}{\circle*{7}}
\multiput(140,140)(40,0){7}{\circle*{7}}
\multiput(180,180)(40,0){6}{\circle*{7}}
\put(20,180){\vector(0,-1){60}}
\put(20,180){\vector(1,0){60}}
\put(85,185){m}
\put(25,115){n}
\put(168,190){(0,0)}
\put(208,190){(0,1)}
\put(248,190){(0,2)}
\put(288,190){(0,3)}
\put(328,190){(0,4)}
\put(368,190){(0,5)}
\put(128,150){(1,-1)}
\put(168,150){(1,0)}
\put(208,150){(1,1)}
\put(248,150){(1,2)}
\put(288,150){(1,3)}
\put(328,150){(1,4)}
\put(368,150){(1,5)}
\put(88,110){(2,-2)}
\put(128,110){(2,-1)}
\put(168,110){(2,0)}
\put(208,110){(2,1)}
\put(248,110){(2,2)}
\put(288,110){(2,3)}
\put(328,110){(2,4)}
\put(368,110){(2,5)}
\put(48,70){(3,-3)}
\put(88,70){(3,-2)}
\put(128,70){(3,-1)}
\put(248,70){(3,2)}
\put(288,70){(3,3)}
\put(168,70){(3,0)}
\put(208,70){(3,1)}
\put(328,70){(3,4)}
\put(368,70){(3,5)}
\put(8,30){(4,-4)}
\put(48,30){(4,-3)}
\put(88,30){(4,-2)}
\put(128,30){(4,-1)}
\put(168,30){(4,0)}
\put(208,30){(4,1)}
\put(248,30){(4,2)}
\put(288,30){(4,3)}
\put(328,30){(4,4)}
\put(368,30){(4,5)}
\end{picture}
\caption{The set $I$}
\end{figure}
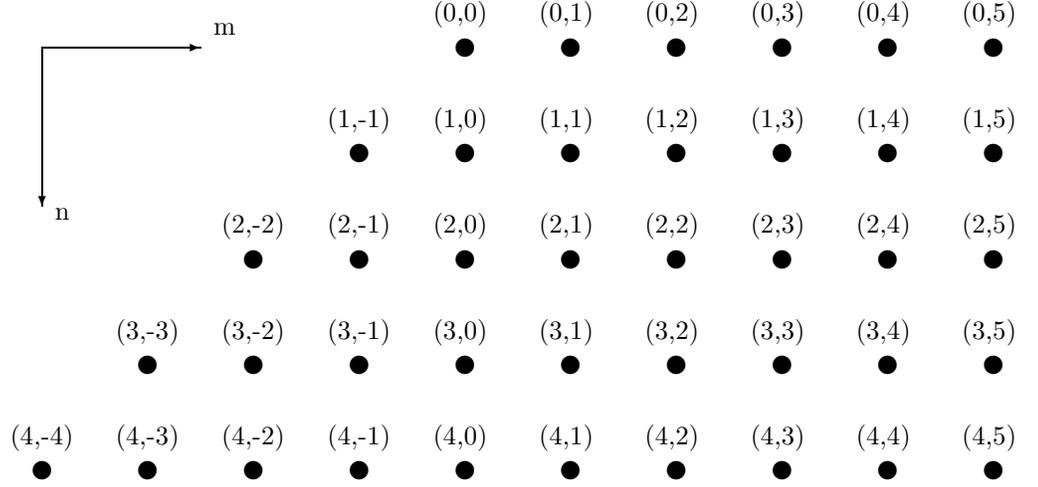
Explicit summation over the $p$'s then leads to a formula for the generating
function of the dimensions of the graded subspaces (which is the character, up
to an overall factor) as
\[ \chi(q,y) \equiv \sum_{n,m} \Dim V_{n,m} \; q^ny^m =\frac{1}{1-y}
\prod_{n=1}^{+\infty} \frac{1}{(1-q^n)(1-q^ny)(1-q^ny^{-1})}\]
Highest weight cyclic modules are quotients of Verma modules.
Thus they are doubly graded,
and we shall see below that their characters are alternating sums of characters
of Verma modules.

\subsection{Singular vectors,
the contragredient form and representation theory}

The first question we have to understand,
now that we have defined Verma modules,
is whether $V^{(\jmath,t)}$ is irreducible as an $U(\hat{\cal A})$-module or
not.
We are going to introduce two important tools that allow us to reformulate this
question and that will also prove useful later on when we discuss fusion:
\begin{itemize}
\item Vectors lying in $V^{(\jmath,t)}$ but not in $V_{0,0}$ and annihilated by
${\cal E}_+$,
called singular vectors,
\item A bilinear symmetric form on $V^{(\jmath,t)}$ called the contragredient
form.
\end{itemize}
We begin by recalling an elementary lemma in linear algebra.

\begin{lem}If a linear map $l$ on a vector space $E$ is diagonalizable,
and if $F$ is a subspace such that $l(F) \subset F$,
then the restriction of $l$ to $F$ is also diagonalizable. \end{lem}
By hypothesis, $E$ is the algebraic direct sum of invariant subspaces of $l$,
that is $E=\oplus_{\lambda} E_{\lambda}$,
with $l_{|E_{\lambda}} =\lambda Id$.
Any vector in $E$,
hence in $F$ is a finite linear combination of eigenvectors with distinct
eigenvalues,
say $f=\sum_{i=1}^n f_i$ with $f_i \in  E_{\lambda_i}$.
By assumption,
$l(F) \subset F$,
hence $f$,
$l(f)$,
$\cdots$,
$l^{(n-1)}(f)$ belong to $F$,
and by inverting the linear system with non-vanishing (Vandermonde) determinant
$\sum_{i=1}^n \lambda_i^j f_i = l^{(j)}(f)$ for $j=0,\cdots,n-1$ we see that
$f_i \in F$ for $i=1,\cdots,n$ and $F=\oplus_{\lambda} E_{\lambda} \bigcap
F$.$\Box$

\vspace{7 mm}

As a consequence of this lemma,
using the operators $\underline{D}$ and $J^0_0$,
we check that any submodule $M$ of  $V^{(\jmath,t)}$ can be decomposed as \[
M=\bigoplus_{(n,m) \in I}  M_{n,m}\]
with $M_{n,m}=V_{n,m} \bigcap M$.
We see here that enlarging ${\cal A}$ in $\hat{\cal A}$ is very useful.
If we define,
for $p \geq 0$,
$M_p=\bigoplus_{2n+m=p} M_{n,m}$,
we end up with a decomposition of $M$ according to the principal gradation.
By the definition of a Verma module if $M_0$ is non-trivial then $M$ coincides
with $V^{(\jmath,t)}$.
If $M$ is a proper submodule we choose $p$ minimal among those for which $M_p
\neq \{0\}$. Then $M_p$ is annihilated by ${\cal E}_+$,
i.e. consists of singular vectors.
Hence any proper submodule of $V^{(\jmath,t)}$ contains a singular vector.
The converse is also true.
In fact let $S'$ be the subspace of $V^{(\jmath,t)}$ annihilated by ${\cal
E}_+$. As $ad(\underline{D})$ and $ad(J^0_0)$ act diagonally on $\hat{\cal A}$,
they map $S'$ into $S'$ and we can apply the lemma  to get
\[ S'=\bigoplus_{(n,m) \in I}  S_{n,m}\]
with $S_{n,m}=V_{n,m} \bigcap S'$. We call $S$ the direct sum of the
$S_{n,m}$'s with $S_{0,0}$ omitted.
If $S \neq  \{0\}$  let $M_S$ be the smallest submodule containing it.
By definition it is $U(\hat{\cal A}) S$ but because $S$ consists of singular
vectors,
by applying the Poincar\'e-Birkhoff-Witt theorem,
this is the same as $U({\cal E}_-) S$. Hence $M_S$ has no intersection with
$V_{0,0}$ and is proper.
But as we saw any proper submodule contains a singular vector,
so is contained in $M_S$.
This proves that $V^{(\jmath,t)}$ either contains no singular vector and is
irreducible,
or contains a unique maximal proper submodule,
generated by the space of singular vectors.
We remarked at the beginning of this section that $U({\cal E}_-)$ does not
contain divisors of zero.
As a consequence we see that a non-zero vector in $S_{n,m}$,
if there is one,
generates under the action of $U(\hat{\cal A})$ (which acts non-trivially only
through $U({\cal E}_-)$) a submodule of $V^{(\jmath,t)}$ isomorphic to
$V^{(\jmath-m,t)}$.
This proves that the characters of cyclic highest weight representations are
alternating sums of characters of Verma modules, as claimed at the end of
section \ref{sec:Vm}.

\vspace{7 mm}

We are now going to recover $M_S$ from another object,
the contragredient bilinear form on $V^{(\jmath,t)}$.
We endow the algebra $\hat{\cal A}$ with the linear anti-automorphism $\sigma$
of order two defined by
\[ \sigma (J^a_n)=J^{-a}_{-n} \qquad \sigma (k)= k \qquad \sigma
(\underline{D}) = \underline{D} \]
As usual this extends in a unique way to a linear anti-automorphism of
$U(\hat{\cal A})$ which we also denote by $\sigma$.
Now to an element $x$ in  $U(\hat{\cal A})$ we associate a complex number
$l(x)$ in the following way.
As $V_{0,0}$ is one dimensional $End(V_{0,0})$ is canonically isomorphic to the
field of complex numbers.
We let $x$ act on $V_{0,0}$ and take the projection of the result back on
$V_{0,0}$.
This defines a linear operator mapping $V_{0,0}$ into itself,
the associated complex number we take to be $l(x)$.
It is clear that $l$ is a linear form on $U(\hat{\cal A})$,
which of course depends on $V^{(\jmath,t)}$.
As we remarked above any element in $U(\hat{\cal A})$ can be written as a
linear combination of terms of the form $x_-x_0x_+$ with $x_a \in U({\cal
E}_a)$ for $a \in \{-,0,+\}$ and $l$ acts on these as 0 except when
$x_-=x_+=1$. But on  $U({\cal E}_0)$,
$\sigma$ acts as the identity.
This proves that $l \circ \sigma =l$.
We can now define $b(x,y) = l(\sigma(x)y)$ for $x$,
$y$ $\in U(\hat{\cal A})$.
Using the properties above we check that $b$ is bilinear and symmetric.
Moreover if $y$ annihilates $V_{0,0}$ then $b(x,y)=0$ for any $x$.
Hence $b$ factors through a bilinear symmetric form on $V^{(\jmath,t)}$.
It is clear that subspaces $V_{n,m}$ indexed by different couples in $I$ are
orthogonal .
We use the notation $\langle x|y \rangle$ for this bilinear form called the
contragredient form.
This notation is reminiscent of the vacuum expectation values in quantum field
theory,
and what we did was just a fancy proof that it was possible to construct such
an expectation value by saying which operator is the adjoint of which (just
what $\sigma$ does).
We denote by $|x\rangle^*$ the linear form associating to $|y\rangle$ the
complex number $\langle y|x \rangle$.
It is readily checked that the kernel of this bilinear form is a proper
(because  $\langle 1|1 \rangle = 1$) submodule,
containing all the singular vectors,
i.e. is nothing but the maximal submodule $M_S$.
We have therefore proved

\begin{theo} The following properties are equivalent:
\begin{enumerate}
\item The module $V^{(\jmath,t)}$ is irreducible.
\item The module $V^{(\jmath,t)}$ contains no singular vector.
\item The contragredient form on $V^{(\jmath,t)}$ is non degenerate.
\end{enumerate}
\end{theo}

\subsection{The Sugawara construction} \label{sec:Sc}

The idea that in some quantum field theories,
the energy-momentum tensor is a suitably renormalized bilinear combination of
the currents proved to have many applications in the representation theory of
affine algebras (see for instance \cite{k}).
We shall see  several examples in the rest of this paper.

\vspace{7 mm}

Let us define elements $C_n$ for integral $n$ by the following formul\ae\ :
\[ C_n = \frac{1}{2} \sum_{m=-\infty}^{+\infty} J^+_{n-m}J^-_m + J^-_{n-m}J^+_m
+ 2J^0_{n-m} J^0_m \qquad \mbox{ for } n \neq 0\]
\[ C_0  = \frac{1}{2} (J^+_0J^-_0 + J^-_0J^+_0 + 2 J^0_0 J^0_0) +
\sum_{m=1}^{+\infty} J^+_{-m}J^-_m + J^-_{-m}J^+_m +2 J^0_{-m} J^0_m \]
A priori these operators live in some completion (to allow infinite sums) of
$U({\cal A})$.
The expression for $C_0$ is some normal ordered version of the generic
expression.
It is easy to see that acting on a state in $V^{(\jmath,t)}$ all but a finite
number of terms in the expression for $C_n$ give 0.
Thus the $C_n$'s are well-defined linear operators on $V^{(\jmath,t)}$.
As such it is well known that they satisfy the following commutation relations
\[ [C_m,J^a_n]=-tnJ^a_{m+n} \qquad  [\underline{D},C_n]=-nC_n \]
\[[C_m,C_n]=t(m-n)C_{m+n} + \frac{1}{4}t(t-2)(m^3-m)\delta_{m+n} \]
So,
for $t \neq 0$,
$V^{(\jmath,t)}$ carries automatically a representation of the Virasoro algebra
with central charge $c=3(t-2)/t$ and conformal weight
$h_{\jmath}=\jmath(\jmath+1)/t$. We set $L_n=C_n/t$.
This leads to
\begin{equation} \label{eq:VKM} [L_m,J^a_n]=-nJ^a_{m+n} \qquad
[\underline{D},L_n]=-nL_n  \end{equation}
\begin{equation} \label{eq:VV} [L_m,L_n]=(m-n)L_{m+n} +
\frac{m^3-m}{12}(3-6t^{-1})\delta_{m+n}\end{equation}
As a byproduct,
we remark that the enlargement of ${\cal A}$ in $\hat{\cal A}$ is also
automatic in the class of representations we are studying.
We simply use $L_0$ instead of $\underline{D}$.
In the next sections we shall need the following expressions for $C_0$ and
$C_{-1}$ which are direct consequences of the definition,
and show clearly their action on the space $S$.
\[ C_0= J^0_0(J^0_0+1) + J^-_0J^+_0 + \sum_{m=1}^{+\infty} J^+_{-m}J^-_m +
J^-_{-m}J^+_m + 2J^0_{-m} J^0_m \]
\[ C_{-1} = J^+_{-1}J^-_0 +2J^0_{-1}J^0_0 +J^-_{-1}J^+_0 + \sum_{m=1}^{+\infty}
J^+_{-m-1}J^-_m + J^-_{-m-1}J^+_m +2 J^0_{-m-1} J^0_m \]

\section{Fundamental results} \label{sec:fr}

We introduce some notations.
The set of couples $(n,m)\in I$ such that $m \neq 0$ and $n$ is a multiple of
$m$ is denoted by $I^{(sing)}$ (see figure~2).
The elements in $I^{(sing)}$ are in one to one correspondence with the elements
of the set $J^{(sing)}$ of couples of integers $(\alpha,\beta)$ such that
$\alpha \neq 0$,
$\beta \geq 0$,
and $\alpha +|\alpha|\beta \geq 0$, by the map $(\alpha,\beta) \to
(|\alpha|\beta,\alpha)$.
We shall often use this parametrization of $I^{(sing)}$.
For $(\alpha,\beta)\in J^{(sing)}$, we define $\jmath_{\alpha,\beta}(t)$ to be
the solution of
\[t|\alpha|\beta+\alpha(2\jmath_{\alpha,\beta}(t) +1 -\alpha)=0\]
\begin{figure}
\begin{picture}(400,200)(-55,0)
\multiput(10,10)(20,0){18}{\circle*{5}}
\multiput(30,30)(20,0){17}{\circle*{5}}
\multiput(50,50)(20,0){16}{\circle*{5}}
\multiput(70,70)(20,0){15}{\circle*{5}}
\multiput(90,90)(20,0){14}{\circle*{5}}
\multiput(110,110)(20,0){13}{\circle*{5}}
\multiput(130,130)(20,0){12}{\circle*{5}}
\multiput(150,150)(20,0){11}{\circle*{5}}
\multiput(170,170)(20,0){10}{\circle*{5}}
\multiput(190,190)(20,0){9}{\circle*{5}}
\multiput(210,190)(0,-20){10}{\circle{10}}
\multiput(230,190)(0,-40){5}{\circle{10}}
\multiput(250,190)(0,-60){4}{\circle{10}}
\multiput(270,190)(0,-80){3}{\circle{10}}
\multiput(290,190)(0,-100){2}{\circle{10}}
\multiput(310,190)(0,-120){2}{\circle{10}}
\multiput(330,190)(0,-140){2}{\circle{10}}
\put(350,190){\circle{10}}
\multiput(170,170)(0,-20){9}{\circle{10}}
\multiput(150,150)(0,-40){4}{\circle{10}}
\multiput(130,130)(0,-60){3}{\circle{10}}
\multiput(110,110)(0,-80){2}{\circle{10}}
\put(90,90){\circle{10}}
\put(70,70){\circle{10}}
\put(50,50){\circle{10}}
\put(30,30){\circle{10}}
\put(10,10){\circle{10}}
\put(20,190){\vector(0,-1){60}}
\put(20,190){\vector(1,0){60}}
\put(85,195){m}
\put(25,125){n}
\put(178,198){(0,0)}
\end{picture}
\caption{The subset $I^{(sing)}$ of $I$.}
\end{figure}
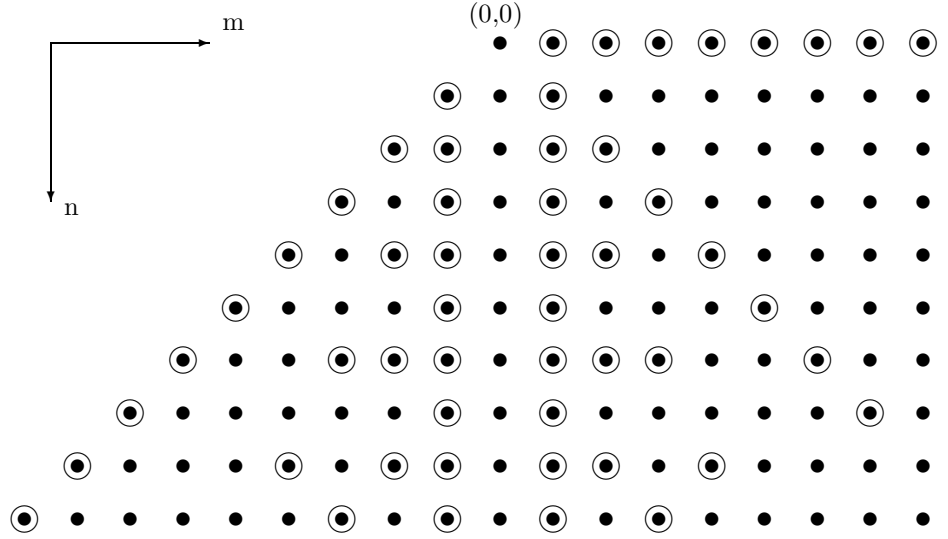

The first theorem,
due to Kac and Kazhdan,
localizes the singular vectors in certain subspaces $V_{n,m}$.

\begin{theo} \label{th:kk}
(Kac-Kazhdan,
\cite{kk}) For nonzero $t$ the Verma module $V^{(\jmath,t)}$ contains a
singular vector at level $(n,m)$ if and only if there is a couple of integers
$(\alpha,\beta)\in J^{(sing)}$ such that $(n,m)=(|\alpha|\beta,\alpha)$ and
$\jmath=\jmath_{\alpha,\beta}(t)$.
Then the dimension of $S_{n,m}$ is exactly one,
i.e. the singular vector is unique up to an overall factor.
\end{theo}

This is an immediate consequence of the following lemma

\begin{lem} (Kac-Kazhdan, \cite{kk}) The determinant $D_{n,m}$ of the
contragredient form in $V_{n,m}$ (defined up to a non-vanishing basis dependent
overall factor) is proportional to
\[ t^{\sum_{\alpha  \geq 1,\beta \geq 1} \Dim   V_{n-\alpha
\beta,m}}\prod_{(\alpha,\beta) \in J^{(sing)} } (t|\alpha|\beta+\alpha(2\jmath
+1 -\alpha))^{\Dim   V_{n-|\alpha|\beta,m-\alpha}} \]
\end{lem}

In the basis (\ref{eq:pbw}), the matrix elements of the contragredient form are
polynomials in $\jmath$ and $t$,
because they are calculated by repeated use of the commutation relations
(\ref{eq:com}).
It follows that $D_{n,m}$ is a polynomial in $t$ and $\jmath$.
To compute this polynomial,
we restrict $t$ and $\jmath$ to be real.
Then there is a basis in which the matrix of the contragredient form is real,
and of course symmetric,
so that it is possible to diagonalize it.
As noted  before,
states with different n or m are orthogonal to each other.

\vspace{7 mm}

First of all we remark that,
because of their commutation relations with the generators of $\hat{\cal A}$,
$C_0$ and $J^0_0$ act diagonally on $V_{n,m}$ with eigenvalues $tn+\jmath
(\jmath+1)$ and $(\jmath-m)$ respectively.
On the other hand,
the explicit expression of $C_0$ shows that $C_0 -J^0_0(J^0_0+1)$ annihilates
the space $S$.
Comparison of these two statements gives: $S_{n,m}$ can be non-trivial only if
$tn+\jmath (\jmath+1) =(\jmath-m) (\jmath-m+1)$ i.e. $tn+m(2\jmath +1 -m) =0$.
This equation plays a very important role in the rest of this paper.

\vspace{7 mm}

We want to study the polynomial $D_{n,m}(\jmath,t)$ as a function of $\jmath$.
For the time being,
we fix $t$ to be some real irrational number.
An elementary computation shows that for fixed $\jmath$ there is at most one
pair $(n_0,m_0) \neq (0,0)$ in $I$ such that $tn_0+m_0(2\jmath +1 -m_0)$ is
zero,
hence at most one non-trivial $S_{n,m}$ in $S$.
 As a further consequence two distinct proper submodules cannot have a
non-trivial intersection,
because it would contain a singular vector in a forbidden place.
But we remarked that a non-trivial vector in $S_{n,m}$ generates a Verma
module.
Hence for $\jmath$ such that $tn_0+m_0(2\jmath +1 -m_0)=0$,
the generating function for the dimensions in $M_S$ is
\[ \chi(q,y) \Dim  (S_{n_0,m_0})q^{n_0}y^{m_ 0}\]
A real symmetric matrix can be diagonalized,
so that when $\jmath$ is such that $tn_0+m_0(2\jmath +1 -m_0)=0$,
the determinant $D_{n,m}$ has a zero of order the dimension of $M_S$ at this
level,
that is $\Dim   S_{n_0,m_0} \Dim   V_{n-n_0,m-m_0}$.
So we end up with a factorization of $D_{n,m}$ as a function of $\jmath$ for
irrational $t$:
\[ D_{n,m} \propto \prod_{(n_0,m_0) \neq (0,0)} (tn_0+m_0(2\jmath +1
-m_0))^{\Dim   S_{n_0,m_0} \Dim   V_{n-n_0,m-m_0}}\]
 The product is in fact finite because $\Dim   V_{n-n_0,m-m_0}$ vanishes for
all but a finite number or couples $(n_0,m_0)$.
The proportionality factor depends on $t$ but in a very simple way.
In fact,
when $t \neq 0$,
if $(n_0,m_0) \neq (0,0)$ in $I$ is such that $tn_0+m_0(2\jmath +1 -m_0)=0$
then $m_0 \neq 0$,
hence if $t \neq 0$ any singular vector gives a factor containing $\jmath$,
and all these have been included.
Thus,
up to a basis dependent constant we have :
\begin{equation} \label{eq:det} D_{n,m} = t^{\sigma_{n,m}}\prod_{(n_0,m_0) \neq
(0,0)} (tn_0+m_0(2\jmath +1 -m_0))^{\Dim   S_{n_0,m_0} \Dim   V_{n-n_0,m-m_0}}
\end{equation}
where the integer $\sigma_{n,m}$ has still to be determined.
The degree in $\jmath$,
which we call $\tau_{n,m}$,
is simply $\sum_{(n_0,m_0) \neq (0,0)} \Dim   S_{n_0,m_0} \Dim
V_{n-n_0,m-m_0}$,
hence the value of the generating function
$\sum_{n,m} \tau_{n,m} q^ny^m$ is
\begin{equation} \label{eq:aa} \chi(q,y) \sum_{n_0,m_0} \Dim   S_{n_0,m_0}
q^{n_0}y^{m_0} \end{equation}

\vspace{7 mm}

We have now going to give an asymptotic estimation of $D_{n,m}$ when we let
$\jmath$ and then $t$ go to infinity.
The trick is simple: we shall change the scales of the generators,
and go to a limit where the affine algebra reduces to an assembly of
independent oscillators.
We define modified generators:
\[ \hat{J}^+_n= \frac{1}{x} J^+_n \qquad \hat{J}^-_n= \frac{1}{x} J^-_n \qquad
\hat{J}^0_n= \frac{1}{y} J^0_n \mbox{ for } n \neq 0 \qquad \hat{J}^0_0=
\frac{1}{x^2} J^0_0 \qquad \hat{k}=\frac{1}{y^2} k\]
We recall that monomials of the form
\[ \prod_{i=1}^{+\infty} (J^+_{-i})^{p_{i,+}} \prod_{i=1}^{+\infty}
(J^0_{-i})^{p_{i,0}} \prod_{i=0}^{+\infty} (J^-_{-i})^{p_{i,-}} |1 \rangle \]
(where all but a finite number of the integers $p$'s are zero) form a basis of
$V^{(\jmath,t)}$.
When we substitute the modified generators for the original ones,
such a monomial picks a factor $x^{\sum p_{i,+} +p_{i,-}}y^{\sum p_{i,0}}$ but
we still have a basis.
It is easy to check that when we let $x$ and then $y$ go to infinity,
the commutation relations for the modified generators have a limiting form
\[
[\hat{J}^+_m,\hat{J}^+_n]=0 \qquad [\hat{J}^-_m,\hat{J}^-_n]=0 \qquad
[\hat{J}^a_m,\hat{k}]=0  \qquad [\hat{J}^0_m,\hat{J}^+_n]= 0 \qquad
[\hat{J}^0_m,\hat{J}^-_n]=0
\]
\[
[\hat{J}^0_m,{J}^0_n]=\frac{\hat{k}}{2}m\delta_{n+m}   \qquad
[\hat{J}^+_m,\hat{J}^-_n]=2 \hat{J}^0_0 \delta_{m+n}
\]
Thus,
in this limit,
the Verma module reduces to a Fock space for independent oscillators,
with $\hat{J}^0_0$ and $\hat{k}$ as normalizations for the scalar products.
The modified monomials form an orthogonal basis for this Fock space.
This scaling argument shows that,
if we let $\jmath$ and then $t$ go to infinity,
the diagonal terms dominate the determinant of the contragredient form,
and contribute in this limit to a factor $(\jmath^{\sum p_{i,+}
+p_{i,-}})(t^{\sum p_{i,0}})$.
Hence every monomial in our basis for the Verma module contributes additively
with a factor $\sum p_{i,+} +p_{i,-}$ to $\tau_{n,m}$ and  $\sum p_{i,0}$ to
$\sigma_{n,m}$ at level $(n,m)=(\sum_{i,a} ip_{i,a},-\sum_{i,a} ap_{i,a})$.
This allows to finish the calculation of the determinant.
Explicit summation over the $p$'s gives
\[ \sum_{n,m} \tau_{n,m} q^ny^m=\chi(q,y) \left( \sum_{i=0}^{\infty}
\sum_{j=1}^{\infty} q^{ij}y^j + \sum_{i=1}^{\infty} \sum_{j=1}^{\infty}
q^{ij}y^{-j} \right)\]
and also
\[\sum_{n,m} \sigma_{n,m} q^ny^m=\chi(q,y)  \sum_{i=1}^{\infty}
\sum_{j=1}^{\infty} q^{ij} \]
Comparison with (\ref{eq:det}) and (\ref{eq:aa}) gives the value of $dim
S_{n_0,m_0}$ and $\sigma_{n,m}$ and leads to the value of the determinant of
the contragredient form
\[D_{n,m}(\jmath,t) \propto  t^{\sum_{\alpha  \geq 1,\beta \geq 1} \Dim
V_{n-\alpha \beta,m}}\prod_{(\alpha,\beta) \in J^{(sing)} }
(t|\alpha|\beta+\alpha(2\jmath +1 -\alpha))^{\Dim
V_{n-|\alpha|\beta,m-\alpha}} \]
where $J^{(sing)}$ is the set of couples of integers $(\alpha,\beta)$ with
$\alpha \neq 0$,
$\beta \geq 0$,
and $\alpha +|\alpha|\beta \geq 0$.$\Box$

For nonzero $t$,
this formula allows the inclusion of Verma modules in  $V^{(\jmath,t)}$ to be
described completely,
leading to explicit formulae for the characters of irreducible cyclic
representations of $A_1^{(1)}$.
The case when $t=0$ is much more complicated (see for instance the conjectures
in \cite{mff}, and the closely related \cite{ff1}).
We shall have very little to say about it in what follows.

\vspace{9 mm}

Now that we know when and where the singular vectors are to be found,
it is possible to look for ``explicit'' expressions.
This was done by Malikov,
Feigin and Fuks in the $A^{(1)}_N$ case.
We quote their result for $N=1$.

\begin{theo} \label{th:mff}
(Malikov-Feigin-Fuks,
\cite{mff}) Fix a nonzero $t$. The vector
\begin{equation} \label{eq:mff1} (J^-_0)^{|\alpha|+t\beta}
(J^+_{-1})^{|\alpha|+t(\beta-1)} (J^-_0)^{|\alpha|+t(\beta-2)}
(J^+_{-1})^{|\alpha|+t(\beta-3)} \cdots (J^-_0)^{|\alpha|-t\beta}
|\jmath_{\alpha,\beta}(t),t
\rangle \end{equation}
for positive $\alpha$ (resp. the vector
\begin{equation} \label{eq:mff2} (J^+_{-1})^{|\alpha|+t(\beta-1)}
(J^-_0)^{|\alpha|+t(\beta-2)}  (J^+_{-1})^{|\alpha|+t(\beta-3)}
(J^-_0)^{|\alpha|+t(\beta-4)}\cdots  (J^+_{-1})^{|\alpha|+t(\beta-1)}
|\jmath_{\alpha,\beta}(t),t\rangle \end{equation}
for negative $\alpha$) is a non-trivial element of $S_{|\alpha| \beta,\alpha}$
in $V^{(\jmath_{\alpha,\beta}(t),t)}$ i.e. is a singular vector.
\end{theo}

These are expressions involving complex exponents of the operators $J^-_0$ and
$J^+_{-1}$,
and they do not make sense a priori.
Malikov,
Feigin and Fuks are able to prove that they make sense by using the following
trick: they prove identities relating products of integral powers of generators
of ${\cal E}_-$, and observe that these identities admit an analytic
continuation for complex powers. Starting from the above expression, by
repeated application of these identities, they end up with a well-defined
expression belonging to $U({\cal E}_-)$ and depending polynomially on $t$.
Moreover,
naive manipulations using the commutation relations as if the exponents where
non-negative integers ``show'' that the above expressions are singular vectors.
Uniqueness of the analytic continuation ensures that this is indeed the case.

In the case when $\alpha$ is a positive integer and $\beta=0$, there is no
analytic continuation to implement, because (\ref{eq:mff1}) reduces to
$(J^-_0)^{\alpha} |\jmath_{\alpha,\beta}(t),t \rangle$. One recovers the
well-known singular vector for the $A_1$-subalgebra $\{J^-_0,J^0_0,J^+_0\}$.
The simplest non-trivial case where analytic continuation is needed is
$(\alpha,\beta)=(1,1)$. We treat this example in appendix \ref{sec:111} to
illustrate the method.

\vspace{7 mm}

It is fair to say that explicit calculations of singular vectors remain quite
complicated,
but these compact formul\ae\ exhibit naturally many non-trivial properties.
Among these, we quote
\begin{itemize}
\item The singular vectors are naturally normalized.
We denoted by ${\cal E}_-$ the Lie algebra of generators of degree (with
respect to the principal gradation $d$) less than 0.
The generators of degree less than $-1$ form an ideal in ${\cal E}_-$,
and we can consider the quotient Lie algebra.
In this quotient $J^-_0$ and $J^+_{-1}$ commute,
and the operators acting on $|\jmath_{\alpha,\beta}(t),t\rangle$ to give the
singular vectors reduce to
$(J^-_0)^{\alpha+|\alpha|\beta}(J^+_{-1})^{|\alpha|\beta}$.
\item Another useful property of the singular vectors is that with the above
normalization they are polynomial in $t$.
\end{itemize}

\vspace{7 mm}

In the rest of this paper we shall give alternative formul{\ae} for the
singular vectors.
They are quite efficient and have an intuitive physical interpretation.
They are connected with fusion rules.
However we have neither been able to show the relation between the two
approaches,
nor to check directly the above properties.

\section{Primary fields and fusion} \label{sec:pff}

We first give some motivation for our abstract definitions,
considering for a while general properties of quantum and conformal field
theories.
Later we shall return to our special case.
In a Euclidean quantum field theory,
we know that short distance singularities in the correlation functions can be
understood in terms of operator product expansions: when the spatial arguments
of two local operators almost coincide,
we can replace their product by some asymptotic expansion in local operators
with functions as coefficients,
and the need to renormalize is responsible for anomalous dimensions.
In 2-d conformal field theory,
the operator product expansion,
also called fusion,
has a much stronger,
and perhaps sounder,
status.
It is  known that its convergence is only limited by the position of the
nearest operator in the correlation function under study.
The symmetries of the theory are rich enough to determine almost completely the
structure of the operator product expansion.
This in turn leads to a purely algebraic or geometric study of the fusion.

\subsection{Motivations}

Any 2-d conformal field theory contains two distinguished operators $T$ and
$\overline{T}$,
which are the components of the traceless symmetric stress-energy tensor in
complex coordinates.
Conservation of stress-energy leads to
\[ \bar{\partial}T=\partial \overline{T}=0 \]
A field $\Phi(w,
\bar{w})$ is called a primary field of weight $(h,\bar{h})$ if its operator
product expansion with $T$ and $\overline{T}$ reads
\begin{equation} \label{eq:Tf} T(z) \Phi(w,
\bar{w})=\left( \frac{h}{(z-w)^2} +\frac{1}{(z-w)}\partial_w \right) \Phi(w,
\bar{w}) \; + \; \mbox{regular terms}\end{equation}
\[ \overline{T}(\bar{z}) \Phi(w,
\bar{w})=\left( \frac{\bar{h}}{(\bar{z}-\bar{w})^2}
+\frac{1}{(\bar{z}-\bar{w})}\bar{\partial}_{\bar{w}} \right) \Phi(w,
\bar{w}) \; + \; \mbox{regular terms}\]
Recalling that the stress-energy tensor generates coordinate transformations,
this simply means that $\Phi(w,
\bar{w})$ is an $(h,\bar{h})$ form in the language of complex geometry.
The fields appearing in this expansion are also scaling fields. They have in
general more singular terms in their short distance expansion with $T$ and
$\overline{T}$.
All the fields one gets by repeated operator product expansions of  $T$ and
$\overline{T}$ with a given primary are called its descendants and they form
what is called a conformal family.
For instance it is a tautology to say that the stress tensor is a descendant of
the identity operator,
and in fact it is not a primary because
\begin{equation} \label{eq:T} T(z)T(w)=\frac{c/2}{(z-w)^4} +
\frac{2T(\frac{z+w}{2})}{(z-w)^2}  \; + \; \mbox{regular terms} \end{equation}
and a similar equation for $\overline{T}$.
This shows that the insertion of $T(z)$ in a correlation function produces a
meromorphic function of $z$,
with known singular part.

\vspace{7 mm}

When one brings two scaling fields $F_1(z,\bar{z})$ and $F_2(w,\bar{w})$ close
together,
one expects that in some weak sense (for instance after insertion in a
correlation function) there is an expansion
\begin{equation} \label{eq:ope} F_1(z,\bar{z})F_2(w,\bar{w})=\sum c_{F_1,F_2}^F
(z-w,\bar{z}-\bar{w}) F(w,\bar{w})\end{equation}
where the sum is over all scaling fields and the coefficients $c_{F_1,F_2}^F$
are functions.
We can split this sum by putting together scaling fields belonging to the same
conformal family.
If (\ref{eq:ope}) is to be true,
both sides of the equality should have the same geometric properties,
i.e. change in the same way under a change of coordinates.
In the field theoretic language,
they should have the same operator product expansion with the components of the
stress-energy tensor (which generates changes of coordinates).
This is only a necessary condition,
but it is very powerful as we shall see.

In the sequel we shall concentrate on the holomorphic part of the conformal
field theory but similar statements hold for the antiholomorphic part.
To go from a formalism of correlation functions to an operator formalism,
we use radial quantization (i.e. decide that the expectation value of a
sequence of operators ordered according to the radial coordinate is simply the
corresponding correlation function) and write $T(z)=\sum_{-\infty}^{+ \infty}
L_n z^{-n-2}$.
A simple application of the Cauchy residue theorem gives an operator
version of (\ref{eq:Tf}) and (\ref{eq:T})
\begin{equation} \label{eq:qTf}[L_m,\Phi(z,\bar{z})] = \left( h(m+1)z^m +
z^{m+1} \partial \right) \Phi(z,\bar{z}) \end{equation}
\begin{equation} \label{eq:qT}[L_m,L_n]=(m-n)L_{m+n} +\frac{c}{12}
(m^3-m)\delta_{n+m} \end{equation}
In particular $L_{-1}$ generates translations and $L_0$ generates dilatations.
If in addition the field $\Phi$ does not depend on $\bar{z}$, it is possible to
expand it as $\Phi(z)=\sum_{-\infty}^{+ \infty} \Phi_n z^{-n-h}$. The structure
of this expansion is dictated by (\ref{eq:qTf}) for $m=0$ and leads to
\begin{equation} \label{eq:qTfc} [L_m,\Phi_n]=(m(h-1)-n)\Phi_{m+n}
\end{equation}

\vspace{7 mm}

Similar considerations apply in the case when holomorphic currents associated
to some semisimple finite dimensional Lie algebra ${\cal G}$ are present.
In this case primary fields have several components.
The translation into operator language of the operator product expansion gives
the commutation relations of the untwisted affine algebra associated to ${\cal
G}$ for the commutators of the currents (that is (\ref{eq:com}) in the
particular case ${\cal G}=A_1$).
For the commutator of a current with a primary field, we get
\begin{equation} \label{eq:qJf}[J^a_n,\Phi_i(z,\bar{z})]=-z^n (R^a)_i^j
\Phi_j(z,\bar{z})\end{equation}
where the matrices $R^a$ carry a representation of ${\cal G}$.
So we see that,
apart from a minus sign,
the commutator acts as the loop algebra in some representation.
Now a descendant  is obtained by repeated operator product expansion of the
currents with a primary.
It should be stressed that although the Sugawara construction leads (in a Verma
module) from the ${\cal G}$ commutation relations to those of the Virasoro
algebra for suitable central charge,
the commutation relations (\ref{eq:qJf}) do not imply that the components of
$\Phi$ are primary fileds for the Sugawara stress tensor.
An explicit calculation shows that one has to postulate the correct commutation
relations with one of the $L_n$'s and then the other follow.
The usual choice is $L_{-1}$,
leading to the Knizhnik--Zamolodchikov equation,
which really is a dynamical equation,
and not a mere tautology.
We see that descendants of a primary field can split into several conformal
families.
By repeated use of these commutation relations (\ref{eq:qTf}) and
(\ref{eq:qJf}) we can evaluate the commutator of any product of primary fields
with the components of the stress-energy tensor or of the currents,
i.e. in a more geometric language the behavior of such a product under a
conformal or a gauge transformation.
For instance if a state $|s\rangle$ is annihilated by some $J^a_n$ then the
state $\Phi_j(z,\bar{z}) |s\rangle$ will be annihilated by $J^a_n\delta^j_i+z^n
(R^a)^j_i$.

If $|\Omega \rangle$ denotes the vacuum state (annihilated by all the $L_n$'s
with $n \geq -1$ and all the $J^a_n$'s with $n \geq 0$ and also their
antiholomorphic partners),
we can create new states by applying a primary field.
The states $\Phi_j(0,0)|\Omega \rangle$ carry a representation of ${\cal G}$
and we can build on this a representation of the associated left and right
affine algebra.
We expect $e^{zL_{-1} +\bar{z}\bar{L}_{-1}}\Phi_j(0,0)|\Omega \rangle$ to
coincide with $\Phi_j(z,\bar{z})|\Omega \rangle$.

\vspace{7 mm}

All these statements made sense in some a priori known conformal field theory,
where operator products were assumed to be well-defined.
Things are quite different when one looks at them from an abstract point of
view.
All one knows from the start is that the space of states should decompose as a
direct sum of representation of the left and right Virasoro algebra or any
larger symmetry algebra ($A_1^{(1)}$  will be the case of interest for us).
No operators are a priori defined,
not mentioning their product.
But,
as we shall see,
the naive manipulations we shall use are close enough to a more axiomatic
approach.
Our construction is completely ``chiral'' in the sense that we completely
forget about antiholomorphic parts.

\subsection{Verma primary fields}

It is time now to return to the $A_1^{(1)}$ case.
We let $t$ be a fixed nonzero complex number (sectors of distinct central
charges are decoupled).

\vspace{7 mm}

First of all we ought to define a vacuum sector.
So we look for a state annihilated  by all the $J^a_n$'s for $n \geq 0$.
This state is to be found in a cyclic module (i.e. a quotient of a Verma
module) and has properties of a highest weight state.
As it should be annihilated by $J^0_0$ the obvious candidate is the highest
weight vector $|0,t\rangle$ in $V^{(0,t)}$.
It not annihilated by $J^-_0$ but clearly $J^-_0|0,t\rangle$ is a singular
vector (according to the theorem \ref{th:kk},
if $t$ is not a rational number,
this is the only singular vector up to normalization),
so we choose for the vacuum sector the resulting quotient and denote the image
of $|0,t\rangle$ (i.e. the vacuum state) by $|\Omega\rangle$.
It is readily checked that if one uses the Sugawara stress tensor the vacuum is
effectively annihilated by the $L_n$'s with $n \geq -1$.

\vspace{7 mm}

We want now to associate a primary field to an arbitrary Verma module
$V^{(\jmath,t)}$.
As we saw above,
the components of this field should carry a representation of the finite
dimensional $A_1$ algebra generated by $J^a_0$,
$a=-,0,+$.
The subspace $\oplus_m V_{(0,m)}$ of $V^{(\jmath,t)}$ carries such a
representation.
It is infinite dimensional,
spanned by the Taylor coefficients  of the family of states
$e^{xJ^-_0}|\jmath,t\rangle$ (parametrized by a complex number $x$).
On this family of states $J^-_0$ acts as $D^-_{\jmath} \equiv
\partial/\partial_x$,
$J^0_0$ as $D^0_{\jmath} \equiv \jmath -x\partial_x$,
and $J^+_0$ as  $D^+_{\jmath} \equiv 2\jmath x -x^2 \partial_x$.
Hence the natural primary field to introduce ought to depend on one extra
variable $x$,
with commutation relations
\begin{equation} \label{eq:comm}  [J^a_n,\Phi_{\jmath}(z,x)]=z^n D^a_{\jmath}
\Phi_{\jmath}(z,x) \end{equation}
We call such a primary field a Verma primary field.
A closely related construction was proposed in \cite{zf}.
This leads to define the action of $\Phi_{\jmath}$ on the vacuum by the formula
\[\Phi_{\jmath}(z,x)|\Omega \rangle \equiv e^{zL_{-1}+xJ^-_0}|\jmath,t\rangle\]
Then we can use repeatedly the commutation relations (\ref{eq:comm}) to define
the action of $\Phi_{\jmath}(z,x)$ on the whole vacuum sector.
For fixed $z$ and $x$,
$e^{zL_{-1}+xJ^-_0}|\jmath,t\rangle$ is not a state in  $V^{(\jmath,t)}$ but
rather in some completion (i.e. in the direct product of the subspaces
$V_{n,m}$) to allow infinite linear combinations.
Of course, if $V^{(\jmath,t)}$ is not irreducible, we can replace it by a
quotient module.

\vspace{7 mm}

Let us mention a more algebraic point of view.
The differential operators ${\cal J}^a_n \equiv -z^n D^a_{\jmath}$  (resp.
${\cal L}_n \equiv -h_{\jmath}(m+1)z^m- z^{m+1} \partial_z$) satisfy formally
the commutation relations of the (non-anomalous) current (resp. Virasoro)
algebra.
Hence the tensor product of $V^{(\jmath,t)}$ with a suitable space of functions
of the variables $x$ and $z$ will carry a graded representation of $A^{(1)}_1$
and of the Virasoro algebra with the correct anomaly.
Thus we can interpret $\Phi_{\jmath}(z,x)|\Omega \rangle$  as an element of
this tensor product having the properties of the vacuum (i.e. it is annihilated
by the same left ideal of $U(\hat{\cal A})$).
We shall see a similar phenomenon when we analyze fusion.

\subsection{Fusion and descent equations} \label{sec:fde}

We shall now try to understand the structure of the operator product expansion
of our Verma primary fields.
Suppose that we bring $\Phi_{\jmath_1}$ and $\Phi_{\jmath_0}$ close together
and look for their operator product expansion.
For our purpose it is sufficient to consider the following state
\begin{equation} \label{eq:p} \Phi_{\jmath_1}(z,x)\Phi_{\jmath_0}(0,0)
|\Omega\rangle \equiv \Phi_{\jmath_1}(z,x)|\jmath_0,t\rangle \end{equation}

\vspace{7 mm}

We postulate the following expansion,
which is the analogue in the operator formalism of the short distance expansion
(\ref{eq:ope})
\begin{equation} \label{eq:op}\Phi_{\jmath_1}(z,x)|\jmath_0,t\rangle
=\sum_{\jmath}|\jmath,t,z,x\rangle \end{equation}
where $|\jmath,t,z,x\rangle$ is a $(z,x)$ dependent state in $V^{(\jmath,t)}$.

Covariance (with respect to the symmetries generated by the current algebra)
implies non trivial constraints for the right hand side of this expansion. This
leads to the following theorem, which is crucial for the rest of our
discussion.

\begin{theo} The covariance of the operator product expansion has the following
consequences:
\begin{enumerate}
\item It fixes the $(z,x)$ dependence of $|\jmath,t,z,x\rangle$ to be
\[|\jmath,t,z,x\rangle=\sum_{(n,m)\in I} z^{h-h_0-h_1+n}
x^{\jmath_0+\jmath_1-\jmath+m} |n,m\rangle_{\jmath}\]
with $|n,m\rangle_{\jmath} \in V_{n,m}$.
\item It leads to relations among the coefficients $|n,m\rangle_{\jmath}$
\begin{equation} \label{eq:-1} J^-_1
|n,m\rangle_{\jmath}=(-\jmath+\jmath_0+\jmath_1+m+1)
|n-1,m+1\rangle_{\jmath}\end{equation}
\begin{equation} \label{eq:+0} J^+_0
|n,m\rangle_{\jmath}=-(-\jmath+\jmath_0-\jmath_1+m-1)
|n,m-1\rangle_{\jmath}\end{equation}
\end{enumerate}
\end{theo}

To find these constraints, we use the following trick: the left ideal in
$U(\hat{\cal A})$ generated by  $J^0_0-\jmath_0$,
$k-(t-2)$,
$L_0-h_0$,
and the $J^a_n$'s in ${\cal E}_+$ annihilates $|\jmath_0,t\rangle$.
Then by using the commutators (\ref{eq:comm}) we get relations that the
right-hand side of (\ref{eq:op}) has to satisfy.
For instance $(J^0_0-\jmath_0)|\jmath_0,t\rangle=0$ implies
$\Phi_{\jmath_1}(z,x) (J^0_0-\jmath_0)|\jmath_0,t\rangle=0$ and after
commutation we get
\begin{equation} \label{eq:m}
(J^0_0-D^0_{\jmath_1}-\jmath_0)\Phi_{\jmath_1}(z,x)|\jmath_0,t\rangle
=0\end{equation}
In the same way we obtain also
\begin{equation} \label{eq:n}(L_0 -h_0 -z\partial_z-h_1)
\Phi_{\jmath_1}(z,x)|\jmath_0,t\rangle =0\end{equation}
and
\[(J^a_n-z^n D^a_{\jmath_1})\Phi_{\jmath_1}(z,x)|\jmath_0,t\rangle
=0  \;\; \forall J^a_p \in {\cal E}_+ \]
As we noticed before,
the corresponding constraints on the right-hand side of (\ref{eq:op}) do not
mix different values of $\jmath$,
and they apply to each term in the sum separately.
So we fix $\jmath$ and decompose $|\jmath,t,z,x\rangle=\sum_{n,m}
|\jmath,t,z,x,n,m\rangle$ according to the eigenvalues of $L_0$ and $J^0_0$.
Then equations (\ref{eq:m}) and (\ref{eq:n}) imply that
\[(\jmath-m-\jmath_0+x\partial_x-\jmath_1)|\jmath,t,z,x,n,m\rangle =0\]
and
\[(h+n-h_0 -z\partial_z-h_1)|\jmath,t,z,x,n,m\rangle =0\]
so they determine completely the $x$ and $z$ dependence.
We write \[|\jmath,t,z,x,n,m\rangle =z^{h-h_0-h_1+n}
x^{\jmath_0+\jmath_1-\jmath+m} |n,m\rangle_{\jmath}\]
with $|n,m\rangle_{\jmath} \in V_{n,m}$.
Then we obtain for the other constraints
\begin{equation} \label{eq:rec-} J^-_p
|n,m\rangle_{\jmath}=(-\jmath+\jmath_0+\jmath_1+m+1) |n-p,m+1 \rangle_{\jmath}
\qquad \mbox{for $p \geq 1$}\end{equation}
\begin{equation} \label{eq:rec0}J^0_p
|n,m\rangle_{\jmath}=-(-\jmath+\jmath_0+m) |n-p,m\rangle_{\jmath} \qquad
\mbox{for $p \geq 1$}\end{equation}
\begin{equation} \label{eq:rec+}J^+_p
|n,m\rangle_{\jmath}=-(-\jmath+\jmath_0-\jmath_1+m-1)
|n-p,m-1\rangle_{\jmath}\qquad \mbox{for $p \geq 0$}\end{equation}
This will be the starting point of the definition of fusion.

We expect that these equations, called the ``descent equations'', are
compatible.
A formal proof of this leads to the definition of a family (parametrized by
$\jmath_0$,
$\jmath_1$ and $\jmath$) of graded representations of ${\cal B}$.
The vector space $V$ on which they act is a direct sum of copies of ${\bf C}$
indexed by couples $(n,m) \in I$,
that is $V=\oplus {\bf C}_{(n,m)}$.
We denote by $\Psi_{n,m}$ the vector with component 1 in ${\bf C}_{(n,m)}$ and
0 elsewhere.
The action of ${\cal B}$ on $V$ is as follows .
The vectors $\Psi_{n,m}$ are eigenvectors $L_0$ and $J^0_0$ with eigenvalue
$h+n$ and $\jmath-m$ respectively. Moreover
\[ J^-_p \Psi_{n,m}=(-\jmath+\jmath_0+\jmath_1+m+1) \Psi_{n-p,m+1} \qquad
\mbox{for $p \geq 1$}\]
\[ J^0_p \Psi_{n,m}=-(-\jmath+\jmath_0+m)  \Psi_{n-p,m}\qquad \mbox{for $p \geq
1$}\]
\[ J^+_p \Psi_{n,m}=-(-\jmath+\jmath_0-\jmath_1+m-1)\Psi_{n-p,m-1} \qquad
\mbox{for $p \geq 0$}\]
Note that we did just mimic the descent equations.
It is easy to check that we indeed get a representation whatever the parameters
$\jmath_0$,
$\jmath_1$ and $\jmath$ are.
We denote these representations by $R^{\jmath}_{\jmath_1,\jmath_0}$. The
representation property implies that the equations
(\ref{eq:rec-},\ref{eq:rec0},\ref{eq:rec+}) are compatible. Then they are
consequences of (\ref{eq:-1},\ref{eq:+0}), because $J^-_1$ and $J^+_0$ generate
${\cal E}_+$ by repeated commutations.$\Box$

We introduce the notation $\mu^a(\jmath-\jmath_0,\jmath_1,m)$ for the scalar
factors on the right hand side of the descent equations,
that is
\[J^a_p \Psi_{n,m}\equiv \mu^a(\jmath-\jmath_0,\jmath_1,m) \Psi_{n-p,m-a} \;\;
\forall (n,m) \in I,
\;\; \forall J^a_p \in {\cal E}_+\]
The striking fact is that $\mu^a(\jmath-\jmath_0,\jmath_1,m)$ does not depend
on the $L_0$ degree.

\vspace{7 mm}

In the formalism of correlation functions, mutually local fields commute. If
they are not mutually local, they do not commute, but after fusion in a given
sector, they commute up to a phase. Thus, in the spirit of radial quantization
we expect that $\Phi_{\jmath_1}(z,x)|\jmath_0,t\rangle$ has exactly the same
covariance properties as (notice the change in the operator ordering)
\begin{equation} \label{eq:pi}
e^{zL_{-1}+xJ^-_0}\Phi_{\jmath_0}(-z,-x)|\jmath_1,t\rangle \end{equation}
We give the proof in appendix \ref{sec:pi}. This property allows these two
states to be identified, as far as covariance is concerned.

\vspace{7 mm}

According to this discussion,
we propose the following definition of fusion.

Fusion of the Verma modules $V^{(\jmath_1,t)}$ and  $V^{(\jmath_0,t)}$ in
$V^{(\jmath,t)}$ is possible if and only if the descent equations
(\ref{eq:rec-},\ref{eq:rec0},\ref{eq:rec+}) have a non-trivial solution. The
dimension of the vector space $E^{\jmath}_{\jmath_1,\jmath_0}$ of solutions of
the set of linear equations (\ref{eq:rec-},\ref{eq:rec0},\ref{eq:rec+}) for the
family of vectors  $|n,m\rangle_{\jmath} \in V_{n,m}$ is called the
multiplicity of the fusion.
A solution of the descent equations is said to be proper if
$|0,0\rangle_{\jmath} \neq 0$.

This deserves some comments.

\begin{itemize}
\item The first point is that we could look for analogous definitions involving
quotient modules of non irreducible Verma modules.
\begin{enumerate}
\item The equations (\ref{eq:rec-},\ref{eq:rec0},\ref{eq:rec+}) still make
sense in any quotient module of  $V^{(\jmath,t)}$ and we can look for solutions
in this smaller space,
modifying the definition of $E^{\jmath}_{\jmath_1,\jmath_0}$ accordingly.
We shall use this generalized definition freely in the following.
\item The case when we consider a quotient module of $V^{(\jmath_1,t)}$ or
$V^{(\jmath_0,t)}$ is more complicated.
We have to introduce new constraints because the ideal annihilating the highest
weight state is bigger.
We shall see examples of this in section \ref{sec:fqvm}.
\end{enumerate}
\item The second point is concerned with the relation between our construction
and the existence of intertwiners between representations.
As we saw above in the definition of Verma primary fields,
the differential operators ${\cal J}^a_n \equiv -z^n D^a_{\jmath_1}$  (resp.
${\cal L}_n \equiv -h_{\jmath_1}(m+1)z^m- z^{m+1} \partial_z$) satisfy formally
the commutation relations of the (non-anomalous) current (resp. Virasoro)
algebra.
Hence the tensor product (denoted by $V^{(\jmath,t)}[z,x]$) of $V^{(\jmath,t)}$
with a suitable space of functions of the variables $x$ and $z$ will carry a
graded representation of $A^{(1)}_1$ and of the Virasoro algebra with the
correct anomaly.
The covariance constraints ensure that the state $\sum_{n,m} z^{h-h_0-h_1+n}
x^{\jmath_0+\jmath_1-\jmath+m} |n,m\rangle_{\jmath}$ associated to a
non-trivial element of $R$ is a highest weight state with highest weight
$\jmath_0$ in this representation.
As such it generates a highest weight module.
Hence there is an intertwiner between  $V^{(\jmath_0,t)}$ and
$V^{(\jmath,t)}[z,x]$.
In the same way one can construct an intertwiner between  $V^{(\jmath_1,t)}$
and $V^{(\jmath,t)}[z,x]$.
Admittedly this is very formal.
We do not attempt to define what we mean by ``suitable space of functions'' and
this prevents us from elucidating the structure the tensor product
representation.
But this suggests that our definition of fusion is reasonably close in spirit
to what is usually done.
Let us also observe that solving the descent equations i.e. finding
$E^{\jmath}_{\jmath_1,\jmath_0}$,
is also an intertwiner problem,
because it amounts to find graded linear maps from
$R^{\jmath}_{\jmath_1,\jmath_0}$ to $V^{(\jmath,t)}$ commuting with the action
of ${\cal B}$.
\item The third point is that we do not impose the absence of short distance
singularities in $x$-space,
that is we do not restrict to the case when $\jmath_1+\jmath_0-\jmath$ is a
nonnegative integer.
This is quite unconventional but well suited to our purposes.
As we shall see in section \ref{sec:fqvm}, when $\jmath_1$ or $\jmath_0$ are
positive integers or half-integers, the singularities in $x$-space disappear.
This is related to the existence of singular vectors (see the first remark
above).
\end{itemize}

Bearing all this in mind,
we can now proceed with the consequences of our definitions.
Let us first explain the content of the descent equations.

\section{First reformulation of the descent equations} \label{sec:frde}

As they stand,
the descent equations are not very tractable.
For given $\jmath_1$,
 $\jmath_0$, and $\jmath$, it is not at all clear whether or not they do have
non-trivial solutions.
However,
we have the following simple bound.

\begin{lem} The vector space of solutions of the descent equations in an
irreducible highest weight cyclic module has dimension at most one.\end{lem}

We introduce a partial ordering on the couples $(n,m)$ by the rule $(n,m) \leq
(n',m')$ if and only if $n \leq n'$ and $n+m \leq n'+m'$ (see example on
figure~3).
With respect to this ordering,
${\cal E}_+$ decreases the degree.
This implies that if the descent equations do have a non-trivial solution in a
highest weight cyclic module,
then the nonzero $|n,m\rangle_{\jmath}$ with minimal $(n,m)$ have to be
annihilated by ${\cal E}_+$.
If the module is irreducible,
the vectors annihilated by ${\cal E}_+$ form a one dimensional subspace
generated by the highest weight state.
Hence any two solutions of the descent equations are proportional.$\Box$
\begin{figure}
\begin{picture}(400,200)(-55,0)
\multiput(10,10)(20,0){18}{\circle*{5}}
\multiput(30,30)(20,0){17}{\circle*{5}}
\multiput(50,50)(20,0){16}{\circle*{5}}
\multiput(70,70)(20,0){15}{\circle*{5}}
\multiput(90,90)(20,0){14}{\circle*{5}}
\multiput(110,110)(20,0){13}{\circle*{5}}
\multiput(130,130)(20,0){12}{\circle*{5}}
\multiput(150,150)(20,0){11}{\circle*{5}}
\multiput(170,170)(20,0){10}{\circle*{5}}
\multiput(190,190)(20,0){9}{\circle*{5}}
\multiput(110,110)(20,0){7}{\circle{10}}
\multiput(130,130)(20,0){7}{\circle{10}}
\multiput(150,150)(20,0){7}{\circle{10}}
\multiput(170,170)(20,0){7}{\circle{10}}
\multiput(190,190)(20,0){7}{\circle{10}}
\put(20,190){\vector(0,-1){60}}
\put(20,190){\vector(1,0){60}}
\put(85,195){m}
\put(25,125){n}
\put(178,198){(0,0)}
\end{picture}
\caption{The couples $(n,m)$ satisfying $(n,m) \leq (4,2)
$}
\end{figure}
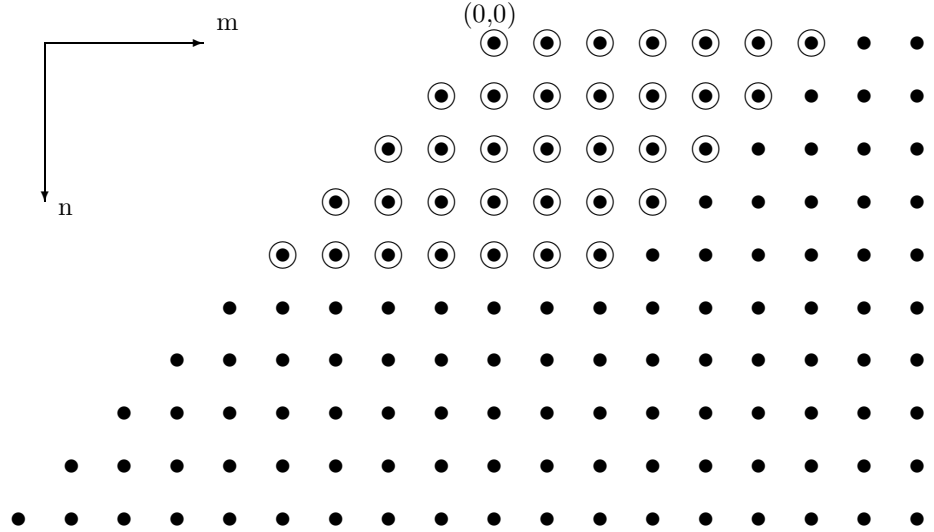

We are going to see that if $V^{(\jmath,t)}$ is irreducible,
the vector space of solutions of the descent equations is exactly one
dimensional.

\vspace{7 mm}

Using the representations $R^{\jmath}_{\jmath_1,\jmath_0}$,
we shall derive consequences of the descent equations which are much easier to
deal with.
We define a family of linear forms on $U({\cal E}_+)$.
Let $x_+$ be in $U({\cal E}_+)$.
If we denote by $\pi$ the projection on ${\bf C}_{(0,0)}$ in
$R^{\jmath}_{\jmath_1,\jmath_0}$,
the composition $\pi x_+$ defines a linear map from ${\bf C}_{(n,m)}$ into
${\bf C}_{(0,0)}$ i.e.
(we identify the endomorphisms of ${\bf C}$ with ${\bf C}$ itself) a complex
number $u_{n,m}(x_+)$,
clearly linear in $x_+$.
Then $u_{n,m} \circ \sigma$ defines a linear form on $U({\cal E}_-)$,
thus on $V^{(\jmath,t)}$.
We denote this form by $\tilde{u}_{n,m}$ and observe that it acts non-trivialy
only on $V_{n,m}$.
As $|0,0\rangle_{\jmath}$ is proportional to the highest weight of
$V^{(\jmath,t)}$,
by applying repeatedly the equations
(\ref{eq:rec-},\ref{eq:rec0},\ref{eq:rec+}) until we end at level $(0,0)$ we do
in fact calculate up to a factor the ``scalar products'' between
$|n,m\rangle_{\jmath}$ and arbitrary elements of $V^{(\jmath,t)}$.
More precisely we have shown
\begin{lem} \label{lem:dual} The descent equations imply that \begin{equation}
\label{eq:dual} |n,m\rangle_{\jmath}^*=\langle \jmath,t|0,0\rangle_{\jmath}
\tilde{u}_{n,m} \qquad \forall (n,m) \in I \end{equation} \end{lem}

If we replace the Verma module $V^{(\jmath,t)}$ by a quotient module,
we have to be careful since $\tilde{u}_{n,m}$ does not always descend to this
quotient.
The obstruction is clearly that $\tilde{u}_{n,m}$ should vanish on the
submodule with respect to which we take the quotient.
However,
the former reasoning shows that if it does not,
the descent equations cannot have a solution in the quotient module.

\subsection{Preliminaries}

To use the full strength of (\ref{eq:dual}),
we need to know some properties of the linear forms $\tilde{u}_{n,m}$.
The action of $\tilde{u}_{n,m}$ on $V^{(\jmath,t)}$ is simple.
We begin with
\begin{lem} \label{lem:van} If $x_-$ is a homogeneous element of $U({\cal
E}_-)$ of degree $(n,m)$,
$\tilde{u}_{n,m}(x_-|\jmath,t \rangle)$ contains a factor $\prod_{i=1}^m
(\jmath-\jmath_0+\jmath_1-m+i)$ if $m>0$ and
$\prod_{i=1}^{-m}(-\jmath+\jmath_0+\jmath_1+m+i)$ if $m<0$.\end{lem}

Without loss of generality,
we can assume that $x_-$ is a monomial in the generators of ${\cal E}_-$.
It is homogeneous in the double gradation,
and we call $(n,m)$ it degree.
We associate to $x_-$ an oriented walk on the set $I$.
The starting point is the pair $(n,m)$.
The operator $\sigma(x_-)$ is a product of generators of ${\cal E}_+$.
Each of these generators defines a step on $I$ according to the double
gradation,
and the walk ends at $(0,0)$.
Knowing the walk allows $x_-$ to be reconstructed.
Relative to the ordering on $I$,
the walk consists of a decreasing sequence.
Now $\sigma(x_-)$ acts on ${\bf C}_{n,m}$  in
$R^{\jmath}_{\jmath_1,\jmath_0}$,
and if our sole purpose is to calculate $\tilde{u}_{n,m}(x_-|\jmath,t
\rangle)$,
we only need to know the projection of the oriented walk on the second factor
(i.e. the space of eigenvalues of $J^0_0$) because the descent equation do not
depend on the projection on the first factor (i.e. the space of eigenvalues of
$L_0$).
This new oriented walk goes from $m$ to 0 and we observe that each step changes
the eigenvalue of $J^0_0$ of at most one unit.
Hence if $m$ is strictly positive,
this walk contains at least once the steps $i \to i-1$ for $i=1,\cdots,m$.
For the same reasons,
if $m$ is strictly negative this walk contains at least once the steps $i \to
i+1$ for $i=m,\cdots,-1$.
This leads to the announced factors.$\Box$

In general,
no other factor is expected,
because there is always at least one monomial $x_-$ whose associated walk
consists (after projection on the second factor) only of decreasing steps if
$m>0$ and increasing steps if $m<0$.

\vspace{7 mm}

To go one step further in the calculation,
we use the particular basis (\ref{eq:pbw}).
Consider the monomial
\[ x_-=\prod_{i=1}^{+\infty} (J^+_{-i})^{p_{i,+}} \prod_{i=1}^{+\infty}
(J^0_{-i})^{p_{i,0}} \prod_{i=0}^{+\infty} (J^-_{-i})^{p_{i,-}} \]
 and set $m_-=\sum_i p_{i,-}$,
$m_0=\sum_i p_{i,0}$,
$m_+=\sum_i p_{i,+}$,
$\sum_{i,a} ip_{i,a}=n$ and $m_--m_+=m$ (then $x_-|\jmath,t \rangle $ belongs
to $V_{n,m}$).
Define polynomials
\[P_{m_-,m_0,m_+}(u,v)=(u-m_-)^{m_0} \prod_{i=1}^{m_+}(v-u+m_--m_++i)
\prod_{i=1}^{m_-}(v+u-m_-+i)\]Then we have
\begin{lem}  The linear form $\tilde{u}_{n',m'}$ takes the value $\delta_{n,n'}
\delta_{m,m'} P_{m_-,m_0,m_+}(\jmath-\jmath_0,\jmath_1)$ on $x_-|\jmath,t
\rangle$ \end{lem}
This is a simple application of the descent equations.$\Box$

As we remarked above,
this ``scalar product'' has no dependence on the $L_0$ gradation,
with the consequence that,
in general,
several monomials $x_-$ lead to the same result.
However,
$(J^+_{-1})^n(J^-_0)^{n+m}$ is the only monomial having $m_-=n+m$ and $m_+=n$.
The next lemma will allow us to prove the existence of fusion rules

\begin{lem} \label{lem:li} For fixed $m$,
the family of polynomials $P_{m_-,m_0,m_--m}$ indexed by $m_-$ and $m_0$ is
linearly independent. \end{lem}
Suppose $\sum_{m_-,m_0} \lambda_{m_-,m_0} P_{m_-,m_0,m_--m}$ is some vanishing
linear combination of these polynomials.
We can group terms to get
\[\sum_{m_-} \left(\sum_{m_0}  \lambda_{m_-,m_0} (u-m_-)^{m_0} \right)
\prod_{i=1}^{m_--m}(v-u+m+i) \prod_{i=1}^{m_-}(v+u-m_-+i) =0 \]
The degree of the polynomials
\[ \prod_{i=1}^{m_--m}(v-u+m+i) \prod_{i=1}^{m_-}(v+u-m_-+i)\]
 in $v$ is $2m_--m$,
thus they are linearly independent as polynomials in $v$.
This implies that $\sum_{m_0}  \lambda_{m_-,m_0} (u-m_-)^{m_0} =0 \;\; \forall
m_-$.
This in turn implies that the initial linear combination was trivial,
i.e. that the $\lambda$'s were all zero.$\Box$

\subsection{Fusion in irreducible Verma modules}

\begin{lem} If  $V^{(\jmath,t)}$ is irreducible,
the vector space of solutions of the descent equations is exactly one
dimensional.
Equivalently,
fusion of $V^{(\jmath_0,t)}$ and $V^{(\jmath_1,t)}$ in an irreducible
$V^{(\jmath,t)}$ is always possible and unique. \end{lem}

In the case when $V^{(\jmath,t)}$ is irreducible,
the contragredient form is non-degenerate.
Hence the equations (\ref{eq:dual}) have a unique solution if we fix the value
of $\langle \jmath,t| 0,0\rangle_{\jmath}$.
This solution is also a solution of the descent equations.
The check is easy.
It is enough to check scalar products.
Let $x_-$ belong to $U({\cal E}_-)$ and $J^a_p$ belong to ${\cal E}_+$.
We have
\begin{eqnarray} \langle x_-|(J^a_p | n,m\rangle_{\jmath})& = & \langle
\sigma(J^a_p)x_-|n,m \rangle_{\jmath} \nonumber \\ & = &
\tilde{u}_{n,m}(\sigma(J^a_p)x_-) \nonumber \\ & = & u_{n,m}(\sigma(x_-)J^a_p)
\nonumber \\ & = & \mu^a(\jmath-\jmath_0,\jmath_1,m)u_{n-p,m-a}(\sigma(x_-))
\nonumber \\ & = & \mu^a(\jmath-\jmath_0,\jmath_1,m)\langle x_-|n-p,m-a
\rangle_{\jmath} \end{eqnarray} $\Box$

In the sequel,
we shall normalize the solution by taking  $|0,0\rangle_{\jmath}=|
\jmath,t\rangle$.

\subsection{Fusion in reducible Verma modules}

In the case when  $V^{(\jmath,t)}$ is reducible,
the contragredient form is degenerate on $M_S$ which is a submodule,
i.e. is stable under the action of $U({\cal E}_-)$.
This implies that the direct sum of the subspaces $V_{n,m}$ on which the
contragredient form is non-degenerate (we call $I'$ the set of couples $(n,m)$
such that this is true, and although $I'$ depends on $\jmath$ and $t$, we shall
not mention this dependence explicitly) is a $U({\cal E}_+)$-module.
Hence the descent equations make sense when restricted to this subspace,
and by the former reasoning,
the vectors $|n,m\rangle_{\jmath}$ for $(n,m) \in I'$ are completely determined
once the value of $\langle \jmath,t| 0,0\rangle_{\jmath}$ has been fixed,
and satisfy the descent equations restricted to this subspace.

\vspace{7 mm}

However,
this solution cannot always be extended to define the states
$|n,m\rangle_{\jmath}$ for $(n,m) \in I \setminus I'$.
This means that fusion rules have made their appearance.
We shall examine them shortly.
They have interest in themselves,
but they will also be of use later on when we shall give formul\ae\ for the
singular vectors.
A word of caution is needed here. For generic values of $t$, there is no hope
of building a respectable conformal field theory, and the word fusion we use
here is an extension of what is usually meant.
\begin{lem}  \label{lem:cf} If $V^{(\jmath,t)}$ is reducible,
fusion is not always possible.
The descent equation have no proper (i.e. such that $|0,0\rangle_{\jmath}\neq
0$) solution in general.
A necessary condition for fusion to be possible is that $\jmath_0$ and
$\jmath_1$ satisfy non-trivial polynomial relations. \end{lem}
We mentioned in section \ref{sec:fr} a crucial property of singular vectors,
called normalization.
We can rephrase it by saying that if $V^{(\jmath,t)}$ contains a singular
vector at level $(n,m)$ (there is no need at this point to be more precise,
but we recall that $(n,m)$ cannot be arbitrary in $I$) and if we expand it in
the basis (\ref{eq:pbw}) the coefficient of $(J^+_{-1})^n(J^-_0)^{n+m}$ is
nonzero and can be rescaled to one (this is the normalization we find if we use
the Malikov,
Feigin and Fuks expressions).
The result on linear independence (lemma \ref{lem:li}) proved in the
preliminaries shows that the value of $\tilde{u}_{n,m}$ on this singular vector
is a non zero polynomial in $\jmath_0$ and $\jmath_1$.
Hence (\ref{eq:dual}) implies that fusion is not possible unless either
$\jmath_0$ and $\jmath_1$ satisfy a non-trivial relation containing $t$ as a
parameter,
or $\langle \jmath,t|0,0\rangle_{\jmath}$ is taken to be zero.$\Box$

These are a priori only necessary conditions.
The second one means that the operator product expansion,
if possible,
is less singular than expected.
Of course,
if $V^{(\jmath,t)}$ contains several singular vectors,
each one contributes a (possibly redundant) constraint on fusion.

If $V^{(\jmath,t)}$ is reducible,
it contains at least one non-trivial submodule,
and we can look for solutions of the descent equations in the quotient module.
As any submodule contains a singular vector,
the proof of the above lemma shows that there is in general an obstruction to
extending the linear forms $\tilde{u}_{n,m}$ to the quotient (see remark after
lemma \ref{lem:dual}),
with the consequence that the fusion rules are also non-trivial in this case.

\subsection{Truncation of the descent equations} \label{sec:Tr}

We shall now see that the descent equation can be truncated in several ways.

\begin{lem} \label{lem:tru} If $-\jmath + \jmath_0 + \jmath_1$ is a nonnegative
integer $i_+$,
it is possible to restrict the descent equations to the subspaces $V_{n,m}$
such that $m \geq -i_+$.
If $-\jmath + \jmath_0 - \jmath_1$ is a nonpositive integer $i_-$,
it is possible to restrict the descent equations to the subspaces $V_{n,m}$
such that $-i_- \geq m$.
\end{lem}

In the first case,
the descent equations connecting the domain $m \geq -i_+$ with the rest of $I$
state that $J^+_p|n,-i_+\rangle_{\jmath}=0$.
In the second case,
the descent equations connecting the domain $-i_- \geq m$ with the rest of $I$
state that $J^-_p|n,-i_-\rangle_{\jmath}=0$.
Hence the announced truncation is possible.$\Box$

\vspace{7 mm}

In fact,
we have a more precise result,
stating that in the rest of $I'$,
the solution of the descent equations is identically $0$.
\begin{lem} If $-\jmath + \jmath_0 + \jmath_1$ is a nonnegative integer $i_+$
and if $(n,m)\in I'$ is such that $m < -i_+$ then $|n,m\rangle_{\jmath}=0$.
If $-\jmath + \jmath_0 -\jmath_1$ is a nonpositive integer $i_-$ and if
$(n,m)\in I'$ is such that $-i_- < m$ then $|n,m\rangle_{\jmath}=0$.
\end{lem}
This is a simple application of lemma \ref{lem:van} and the fact that the
contragredient form is nondegenerate on $V_{n,m}$,
$(n,m)\in I'$.$\Box$

\vspace{7 mm}

If both the above conditions are satisfied,
(in which case $\jmath_1=1/2(i_+-i_-)$ is a nonnegative integer or
half-integer) this truncation is related to the fusion of quotients of Verma
modules. This is shown in appendix \ref{sec:fqvm}, where a derivation, using
our technique, of the (well known) fusion rules for the unitary models is also
given.

\subsection{Algebraic structure of the solutions of the descent equations}
\label{sec:ass}

To close this section,
we make some comments on the behavior of the solutions of the descent equation
as functions of the parameters $\jmath$,
$\jmath_0$,
$\jmath_1$ and $t$.

We already remarked that all Verma modules are isomorphic  to $U({\cal E}_-)$
as $U({\cal E}_-)$-modules.
This allows us to consider them in a uniform way.
\begin{lem} \label{lem:pol} The action of $\hat{\cal A}$ (hence of $U(\hat{\cal
A})$) on $V^{(\jmath,t)}$ is polynomial in $\jmath$ and $t$.\end{lem}
To give a precise content to this lemma,
we use our preferred basis (\ref{eq:pbw}) in $V^{(\jmath,t)}$ to write down the
matrices of the linear maps $J^a_p$ mapping $V_{n,m}$ into $V_{n-p,m-a}$.
That the matrix elements are polynomial in $\jmath$ and $t$ (in fact of degree
$\leq 1$) is an immediate consequence of the constitutive commutation relations
(\ref{eq:com}).
The same property of course holds if we choose another $\jmath$ and $t$
independent basis of  $U({\cal E}_-)$.$\Box$

\vspace{7 mm}

If $V^{(\jmath,t)}$ is irreducible,
we have seen that the descent equations have exactly one normalized solution
and we can interpret the normalized sequence $|n,m \rangle_{\jmath}$ as a
sequence $x_-^{n,m}$ in  $U({\cal E}_-)$.
\begin{lem} Each $x_-^{n,m}$ is rational in $\jmath$ and $t$ and polynomial in
$\jmath_0$ and $\jmath_1$.
The poles in $\jmath$ can occur only at zeroes $\jmath_{\alpha,\beta}(t)$ of
the determinant of the contragredient form.
 \end{lem}
According to theorem \ref{th:kk},
for fixed $(n,m)$ and $t\neq 0$,
there is only a finite number of values of $\jmath$ such that the
contragredient form is degenerate on $V_{n,m}$ in $V^{(\jmath,t)}$.
The determinant of the contragredient form is polynomial in $\jmath$ and $t$
and the linear forms $\tilde{u}_{n,m}$ evaluated at members of the basis
(\ref{eq:pbw}) depend on  $\jmath$,
$\jmath_0$ and $\jmath_1$ polynomially.
Hence the solution of the system (\ref{eq:dual}),
whose determinant is the determinant of the contragredient form at level
$(n,m)$,
has the announced properties.$\Box$

\vspace{7 mm}

The singularities of $x_-^{n,m}$ as a function of $\jmath$ and $t$ may depend
on the value of $\jmath_0$ and $\jmath_1$.
The two above lemmas lead to the following
\begin{cor} \label{cor:ac1} If for a certain choice of $(\alpha,\beta)$,
$\jmath_0$ and $\jmath_1$,
each and every $x_-^{n,m}$ has a limit when $\jmath$ goes to
$\jmath_{\alpha,\beta}(t)$,
then the image of the limit of $x_-^{n,m}$ in
$V^{(\jmath_{\alpha,\beta}(t),t)}$ gives a solution of the descent equations.
\end{cor}

%

\section{Second reformulation of the descent equations} \label{sec:srde}

We are now going to derive the most useful consequences of the descent
equations.
Then,
we shall give a geometric interpretation to our computations.

\subsection{Triangular form of the descent equations}

The fundamental result is

\begin{lem} \label{lem:tri} Any solution of the descent equations satisfies
\begin{eqnarray} \label{eq:tri} \left(tn+m(2\jmath +1
-m)\right)|n,m\rangle_{\jmath}&=&  (-\jmath+\jmath_0+\jmath_1+m+1)
\sum_{p=1}^n J^+_{-p}|n-p,m+1 \rangle_{\jmath} \nonumber \\ & &
-2(-\jmath+\jmath_0+m)\sum_{p=1}^n J^0_{-p} |n-p,m\rangle_{\jmath} \\ & &
 -(-\jmath+\jmath_0-\jmath_1+m-1)\sum_{p=0}^n J^-_{-p}|n-p,m-1\rangle_{\jmath}
\nonumber \end{eqnarray} for $(n,m) \neq (0,0)$ \end{lem}

Multiply the descent equations (\ref{eq:rec-}),
(\ref{eq:rec0}) and (\ref{eq:rec+}) by $J^+_{-p}$,
$J^0_{-p}$ and $J^-_{-p}$ respectively.
Then the sum $\sum_{p=1}^n J^+_{-p}(\ref{eq:rec-}) +2 \sum_{p=1}^n
J^0_{-p}(\ref{eq:rec0})+\sum_{p=0}^n J^-_{-p}(\ref{eq:rec+})$ gives on the
right hand side of the equality the right hand side of (\ref{eq:tri}).
On the left hand side,
one recognizes the definition of $\left( C_0-J^0_0(J^0_0+1) \right)
|n,m\rangle_{\jmath}$,
which is nothing but the left hand side of (\ref{eq:tri}).$\Box$

For $(n,m)=(0,0)$ it is natural to interpret (\ref{eq:tri}) as the empty
relation $0=0$.

\vspace{7 mm}

It will be useful later on to separate the equation (\ref{eq:tri}) to get a
system
\[\left\{ \begin{array}{rcl} \overline {|n,m\rangle_{\jmath}}&=&
(-\jmath+\jmath_0+\jmath_1+m+1)  \sum_{p=1}^n J^+_{-p}|n-p,m+1 \rangle_{\jmath}
 \\ & &  -2(-\jmath+\jmath_0+m)\sum_{p=1}^n J^0_{-p} |n-p,m\rangle_{\jmath}  \\
& &
 -(-\jmath+\jmath_0-\jmath_1+m-1)\sum_{p=0}^n
J^-_{-p}|n-p,m-1\rangle_{\jmath}\\ \left(tn+m(2\jmath +1
-m)\right)|n,m\rangle_{\jmath}&=& \overline{|n,m\rangle}_{\jmath} \end{array}
\right.\]

\vspace{7 mm}

The important property of equation (\ref{eq:tri}) is its triangular structure.
The appearance of the prefactor $tn+m(2\jmath +1 -m)$ should not come as a
surprise.
If this prefactor does not vanish,
the state $|n,m\rangle_{\jmath}$ is expressed in terms of states of lower
degree (we still use the same ordering in $I$).
Hence,
if $\jmath$ and $t$ are such that $tn+m(2\jmath +1 -m)$ vanishes for no
non-trivial value of $(n,m)$ (this is more restrictive than demanding that
$\jmath$ is not a $\jmath_{\alpha,\beta}(t)$),
(\ref{eq:tri}) has a unique proper normalized solution,
whatever the values of $\jmath_0$ and $\jmath_1$ are.
By unicity,
this solution has to be a solution of the descent equations.
However,
we can show a little more.

\vspace{7 mm}

For fixed values of $\jmath$ and $t$,
we call $I''$ the subset of $I$ containing the set of pairs $(n',m')$ such that
$tn+m(2\jmath +1 -m)\neq 0$ for any $(n,m)\in I \setminus (0,0)$ such that
$(n,m)\leq (n',m')$.
The set $I''$ contains $(0,0)$.

\begin{lem} \label{lem:us}The equation (\ref{eq:tri}) restricted to $I''$ has a
unique normalized solution,
and this solution satisfies the descent equations.\end{lem}
By the definition of $I''$,
the direct sum $\oplus_{(n,m)\in I''} V_{n,m}$ is a $U({\cal E}_+)$-module.
Hence the equation (\ref{eq:tri}) and the descent equations make sense when
restricted to this subspace of $V^{(\jmath,t)}$.
It is clear from the triangular structure of (\ref{eq:tri}) that the restricted
equation has a unique solution.
As $I''$ is included in $I'$,
we know that the descent equations also have a unique normalized solution for
$(n,m)\in I''$.
These solutions have to coincide.$\Box$

\vspace{7 mm}

We also have a weaker result when $(n,m)$ is ``as close as possible'' to
$I''$.
\begin{lem} \label{lem:eus} Let $(n,m)\in I$ be such that $(n',m') < (n,m)$
implies $(n',m') \in I''$.
Then
\[J^a_p \overline{|n,m\rangle}_{\jmath}=\left(tn+m(2\jmath +1 -m)\right)
\mu^a(\jmath-\jmath_0,\jmath_1,m) |n-p,m-a\rangle_{\jmath} \;\; \forall J^a_p
\in {\cal E}_+ \]
\end{lem}
Let us first note that,
with the hypotheses of the lemma, either $(n,m)$ belongs to $I''$ or
$tn+m(2\jmath +1 -m)=0$. According to lemma \ref{lem:us},
$\overline{|n,m\rangle}_{\jmath}$,
which is expressed only in terms of vectors $|n',m'\rangle_{\jmath}$ with
$(n',m') \in I''$,
is well-defined.
To prove the lemma,
it is enough to check the cases $J^a_p=J^+_0$ and  $J^a_p=J^-_1$.
We do the calculation in detail for $J^+_0$,
and leave the other verification to the motivated reader. Using the commutation
relations (\ref{eq:com}) we obtain
\begin{eqnarray} J^+_0 \overline{|n,m\rangle}_{\jmath}&=&
(-\jmath+\jmath_0+\jmath_1+m+1)  \sum_{p=1}^n J^+_{-p}J^+_0|n-p,m+1
\rangle_{\jmath} \nonumber \\ & &  -2(-\jmath+\jmath_0+m)\sum_{p=1}^n
(J^0_{-p}J^+_0-J^+_{-p}) |n-p,m\rangle_{\jmath} \nonumber \\ & &
 -(-\jmath+\jmath_0-\jmath_1+m-1)\sum_{p=0}^n (
J^-_{-p}J^+_0+J^0_{-p})|n-p,m-1\rangle_{\jmath}\end{eqnarray}

On the right hand side,
the descent equations are valid,
because we can simply invoke lemma \ref{lem:us} (Notice that we might also
argue by induction as follows.
The vector $|0,0\rangle_{\jmath}$ always satisfies the descent equations.
We assume that the descent equations are valid for the predecessors of $(n,m)$
and we follow the rest of the proof of lemma \ref{lem:eus}.
Then if $(n,m)$ belongs to $I''$,
$tn+m(2\jmath +1 -m)$ does not vanish and we infer that $|n,m\rangle_{\jmath}$
is well-defined and satisfies the descent equations,
completing the induction step and giving an alternative proof of \ref{lem:us}).
Using the descent equations we get
\begin{eqnarray} J^+_0 \overline{|n,m\rangle}_{\jmath}&=&
-(-\jmath+\jmath_0+\jmath_1+m+1)(-\jmath+\jmath_0-\jmath_1+m)  \sum_{p=1}^n
J^+_{-p}|n-p,m\rangle_{\jmath} \nonumber \\ & &
+2(-\jmath+\jmath_0+m)(-\jmath+\jmath_0-\jmath_1+m-1) \sum_{p=1}^n
J^0_{-p}|n-p,m-1\rangle_{\jmath}  \nonumber \\ & &+2(-\jmath+\jmath_0+m)
\sum_{p=1}^n J^+_{-p} |n-p,m\rangle_{\jmath} \nonumber \\ & &
 +(-\jmath+\jmath_0-\jmath_1+m-1)(-\jmath+\jmath_0-\jmath_1+m-2)\sum_{p=0}^n
J^-_{-p}|n-p,m-2\rangle_{\jmath} \nonumber \\ & &
-(-\jmath+\jmath_0-\jmath_1+m-1) \sum_{p=0}^n
J^0_{-p}|n-p,m-1\rangle_{\jmath}\end{eqnarray}

We recognize many terms of the right hand side of (\ref{eq:tri}) for the couple
$(n,m-1)$.
We obtain
\begin{eqnarray} J^+_0 \overline{|n,m\rangle}_{\jmath}&=&
-(-\jmath+\jmath_0-\jmath_1+m-1)(tn+(m-1)(2\jmath+2-m)) |n,m-1\rangle_{\jmath}
\nonumber \\ & & -2(-\jmath+\jmath_0+m) \sum_{p=1}^n
J^+_{-p}|n-p,m\rangle_{\jmath} \nonumber \\ & &
+2(-\jmath+\jmath_0-\jmath_1+m-1) \sum_{p=1}^n J^0_{-p}|n-p,m-1\rangle_{\jmath}
 \nonumber \\ & &+2(-\jmath+\jmath_0+m)  \sum_{p=1}^n J^+_{-p}
|n-p,m\rangle_{\jmath} \nonumber \\ & &
-2(-\jmath+\jmath_0-\jmath_1+m-1) \sum_{p=0}^n J^0_{-p}|n-p,m-1\rangle_{\jmath}
\end{eqnarray}

There are many cancellations on the right hand side,
and except for the first line and the term $p=0$ in the last line,
everything disappears.
But $J^0_0$ acts on $|n-p,m-1\rangle_{\jmath}$ as multiplication by
$\jmath-m+1$,
and we finally obtain
\[J^+_0 \overline{|n,m\rangle}_{\jmath}=
-(-\jmath+\jmath_0-\jmath_1+m-1)(tn+m(2\jmath+1-m))
|n,m-1\rangle_{\jmath}\]$\Box$

\vspace{7 mm}

We deduce the following result, which is reminiscent of corollary
\ref{cor:ac1}.
For fixed nonzero $t$, we can consider the solution of the equation
(\ref{eq:tri}) as a function of $\jmath$, $\jmath_0$ and $\jmath_1$.
A given couple $(n,m)$ belongs to $I''$ for all but a finite number of values
of $\jmath$,
and the form of equation (\ref{eq:tri}) gives another proof that the vectors
$x^{n,m}_-\in U({\cal E_-})$, introduced in section \ref{sec:ass}, are rational
in $\jmath$ and $t$ and polynomial in $\jmath_0$ and $\jmath_1$. However the
prefactor $tn+m(2\jmath+1-m)$ in (\ref{eq:tri}) leads to consider ``spurious''
poles for $x^{n,m}$. We know that the true poles are the zeroes of the
determinant of the contragredient form. Hence, the only couples $(n,m)$ that
contribute to the poles are of the form $(|\alpha|\beta,\alpha)$ for
$(\alpha,\beta)\in J^{(sing)}$.
\begin{cor} \label{cor:ac2}  Let $(n,m)\in I$ be such that for
$\jmath=-t\frac{n}{2m} +\frac{m-1}{2}$, $(n',m') < (n,m)$ implies $(n',m') \in
I''$.
If $\overline{|n,m\rangle}_{-t\frac{n}{2m} +\frac{m-1}{2}} =0$, then
$x^{n,m}_-$ has a limit when $\jmath \to -t\frac{n}{2m} +\frac{m-1}{2}$. The
image of this limit in $V^{(-t\frac{n}{2m} +\frac{m-1}{2},t)}$ satisfies the
descent equations at degree $(n,m)$. \end{cor}
We are interested in the behavior of $|n,m\rangle_{\jmath}$ near
$\jmath=-t\frac{n}{2m} +\frac{m-1}{2}$.
The vector $\bar{x}^{n,m}_-\in U({\cal E_-})$ (corresponding to
$\overline{|n,m\rangle}_{\jmath} \in V^{(\jmath,t)}$) is well-defined and
analytic in $\jmath$ in a neighborhood of $-t\frac{n}{2m} +\frac{m-1}{2}$.
Hence the vanishing of $\overline{|n,m\rangle}_{-t\frac{n}{2m} +\frac{m-1}{2}}$
implies that $(tn+m(2\jmath+1-m))^{-1} \bar{x}^{n,m}_-$ has a limit when
$\jmath \to -t\frac{n}{2m} +\frac{m-1}{2}$.
We take this limit to be $x^{n,m}_-$ at the point $\jmath=-t\frac{n}{2m}
+\frac{m-1}{2}$.
The proof that this limit satisfies the descent equations at degree $(n,m)$ is
the same as the proof of corollary \ref{cor:ac1}.$\Box$

\subsection{The Knizhnik--Zamolodchikov equation}

Although the derivation of (\ref{eq:tri}) is simple,
its physical meaning is not clear.
We shall now show that (\ref{eq:tri}) is a consequence of the
Knizhnik--Zamolodchikov equation,
illuminating the geometrical origin of the descent equations and their
associated triangular form.

\begin{lem} \label{lem:KZ} Equation (\ref{eq:tri}) is the constraint on
$\Phi_{\jmath_1}(z,x)|\jmath_0,t\rangle$ coming from the fact that $tL_{-1}-
J^+_{-1}J^-_0 -2J^0_{-1}J^0_0 $  annihilates the state $|\jmath_1,t\rangle$.
\end{lem}
The proof is a straightforward but tedious computation.
Remark that  $(tL_{-1}- J^+_{-1}J^-_0 -2J^0_{-1}J^0_0)|\jmath_1,t\rangle=0$
comes from the definition of $L_{-1}$ by the Sugawara construction.
On $|\jmath_1,t\rangle$,
$J_0^0$ acts as multiplication by $\jmath_1$,
so we start with
\[(tL_{-1}- J^+_{-1}J^-_0 -2\jmath_1 J^0_{-1})|\jmath_1,t\rangle=0\]
and multiply on the left by $ e^{zL_{-1}+xJ^-_0}\Phi_{\jmath_0}(-z,-x)$.
We use the commutation relations (\ref{eq:qTf}) and (\ref{eq:comm}) to get
\begin{eqnarray*}
\lefteqn{e^{zL_{-1}+xJ^-_0}
(t(L_{-1}+\partial_z)-(J^+_{-1}-z^{-1}D^+_{\jmath_0})(J^-_0
+D^-_{\jmath_0})-2\jmath_1 (J^0_{-1}+z^{-1}D^0_{\jmath_0}))
e^{-zL_{-1}-xJ^-_0}}\qquad \qquad \qquad \qquad \qquad \qquad \qquad \qquad
\qquad \qquad \\ & &
e^{zL_{-1}+xJ^-_0}\Phi_{\jmath_0}(-z,-x)|\jmath_1,t\rangle=0 \end{eqnarray*}
We have checked in lemma \ref{lem:icc} that, as far as covariance is concerned,
it is not possible to distinguish
$e^{zL_{-1}+xJ^-_0}\Phi_{\jmath_0}(-z,-x)|\jmath_1,t\rangle$ and
$\Phi_{\jmath_1}(z,x)|\jmath_0,t\rangle$.
Hence we have to compute
\begin{equation} \label{eq:KZ'} e^{zL_{-1}+xJ^-_0}
(t(L_{-1}+\partial_z)-(J^+_{-1}-z^{-1}D^+_{\jmath_0})(J^-_0
+D^-_{\jmath_0})-2\jmath_1 (J^0_{-1}+z^{-1}D^0_{\jmath_0}))
e^{-zL_{-1}-xJ^-_0}\end{equation}
This is done by repeated use of the commutation relations (\ref{eq:com}) and
(\ref{eq:VKM}).
We compute
\begin{eqnarray*} e^{zL_{-1}+xJ^-_0} J^+_{-1}e^{-zL_{-1}-xJ^-_0}&= &
e^{zL_{-1}}( J^+_{-1}-2xJ^0_{-1}-x^2J^-_{-1}) e^{-zL_{-1}}\\ & = &
\sum_{p=1}^{\infty} z^{p-1}( J^+_{-p}-2xJ^0_{-p}-x^2J^-_{-p}) \\
e^{zL_{-1}+xJ^-_0} J^0_{-1}e^{-zL_{-1}-xJ^-_0}&= &
e^{zL_{-1}}(J^0_{-1}+xJ^-_{-1}) e^{-zL_{-1}}\\& = & \sum_{p=1}^{\infty} z^{p-1}
(J^0_{-p}+xJ^-_{-p}) \\ e^{zL_{-1}+xJ^-_0} J^-_{0}e^{-zL_{-1}-xJ^-_0}&= &
J^-_{0}  \end{eqnarray*}
We define $\tilde{J}^+(z)=\sum_{p=1}^{\infty} z^{p-1} J^+_{-p}$,
$\tilde{J}^0(z)=\sum_{p=1}^{\infty} z^{p-1} J^0_{-p}$ and
$\tilde{J}^-(z)=\sum_{p=0}^{\infty} z^{p-1} J^+_{-p}$.
We can interpret these expressions as the ``negative part'' of the currents,
the part which acts non-trivially on the highest weight state.
The $p=0$ part of $\tilde{J}^-(z)$ appears in the computation of
\begin{eqnarray*} e^{zL_{-1}+xJ^-_0}D^+_{\jmath_0}e^{-zL_{-1}-xJ^-_0} & = &
D^+_{\jmath_0} + x^2 J^-_0 \\
e^{zL_{-1}+xJ^-_0}D^0_{\jmath_0}e^{-zL_{-1}-xJ^-_0} & = & D^+_{\jmath_0} + x
J^-_0 \end{eqnarray*}
It is now a simple matter of regrouping terms to check that (\ref{eq:KZ'}) is
equal to
\begin{equation} \label{eq:KZ} (t\partial_z +
z^{-1}(D^+_{\jmath_1}D^-_{\jmath_1}-2\jmath_0D^0_{\jmath_1})) -
(\tilde{J}^+(z)D^-_{\jmath_1}+2\tilde{J}^0(z)
D^0_{\jmath_1}+\tilde{J}^-(z)D^+_{\jmath_1}) \end{equation}
The exchange of $\jmath_0$ and $\jmath_1$ is somewhat unexpected,
but in fact $D^+_{\jmath_1}D^-_{\jmath_1}-2\jmath_0D^0_{\jmath_1} =
D^+_{\jmath_0}D^-_{\jmath_0}-2\jmath_1D^0_{\jmath_0}$.
If we apply (\ref{eq:KZ}) to the short distance expansion projected on the
$\jmath$-sector
\[ \sum_{n,m} z^{h-h_0-h_1+n} x^{\jmath_0+\jmath_1-\jmath+m}
|n,m\rangle_{\jmath}\]
we know that we obtain zero.
Term by term identification of the powers of $z$ and $x$ leads to
(\ref{eq:tri}).$\Box$

By abuse of language,
we call (\ref{eq:tri}) the fused  Knizhnik--Zamolodchikov equation.

\vspace{7 mm}

The fundamental role played by the Knizhnik--Zamolodchikov equation, or its
fused version (\ref{eq:tri})), is not really a surprise.
It is well known that this equation is related to the existence of integral
representations (i.e. quite explicit forms) for the correlation functions of
minimal $A^{(1)}_1$ Wess--Zumino--Witten models (see for instance \cite{sv}).
This shows that it is related to the fusion,
but also to the structure of singular vectors.
We shall see shortly that this is indeed true.

\section{Singular vectors} \label{sec:sv}

We are finally in position to propose an effective way to compute singular
vectors.

\subsection{General construction}

We fix a nonzero $t$.
\begin{lem}   Let $(n,m)\in I$ be such that for $\jmath=-t\frac{n}{2m}
+\frac{m-1}{2}$, $(n',m') < (n,m)$ implies $(n',m') \in I''$.
Then $\overline{|n,m\rangle}_{-t\frac{n}{2m} +\frac{m-1}{2}}$ is annihilated by
$U({\cal E}_+)$ \end{lem}
If $(n,m)$ satisfies the hypotheses, $\overline{|n,m\rangle}_{-t\frac{n}{2m}
+\frac{m-1}{2}}$ is well-defined.
The lemma is then a direct consequence of lemma \ref{lem:eus}.$\Box$

\begin{cor} Under the same hypotheses, if
$\overline{|n,m\rangle}_{-t\frac{n}{2m} +\frac{m-1}{2}}$ does not vanish, it is
a singular vector.\end{cor}
Clear from the definition of the singular vector.$\Box$

\begin{cor} Under the same hypotheses, if $(n,m)$ is not of the form
$(|\alpha|\beta,\alpha)$ for some $(\alpha,\beta)\in J^{(sing)}$,
$\overline{|n,m\rangle}_{-t\frac{n}{2m} +\frac{m-1}{2}}$ does vanish. \end{cor}
Clear because in this case  $V^{(-t\frac{n}{2m} +\frac{m-1}{2},t)}$ contains no
singular vector.$\Box$

\begin{lem} Let $(|\alpha|\beta,\alpha)\in I$ be such that for
$\jmath=\jmath_{\alpha,\beta}(t)$, $(n',m') < (|\alpha|\beta,\alpha)$ implies
$(n',m') \in I''$. As a polynomial in $\jmath_0$ and $\jmath_1$,
$\overline{|\beta|\alpha|,\alpha\rangle}_{\jmath_{\alpha,\beta}(t)}$ cannot
vanish identically. \end{lem}
As we have seen in the proof of lemma \ref{lem:cf}, if $V^{(\jmath,t)}$
contains a singular vector at level $(n,m)$, the equation
\begin{equation} \label{eq:dul} |n,m\rangle_{\jmath}^*=\langle
\jmath,t|0,0\rangle_{\jmath} \tilde{u}_{n,m} \end{equation}
cannot have a solution, unless $\jmath_0$ and $\jmath_1$ satisfy non-trivial
relations.
But corollary \ref{cor:ac2} shows that whenever
$\overline{|\beta|\alpha|,\alpha\rangle}_{\jmath_{\alpha,\beta}(t)}$ vanishes
(for a particular value of $\jmath_0$ and $\jmath_1$), it is possible to define
a solution of the descent equations at level $(|\alpha|\beta,\alpha)$ by
analytic continuation.
This solution is automatically a solution of (\ref{eq:dul}).$\Box$

This leads to the important

\begin{theo} \label{theo:fin} Let $t$ be irrational. Unless $\jmath_0$ and
$\jmath_1$ satisfy non-trivial fusion rules, the vector
$\overline{|\beta|\alpha|,\alpha\rangle}_{\jmath_{\alpha,\beta}(t)}$ is a
non-vanishing  singular vector in $V^{(\jmath_{\alpha,\beta}(t),t)}$ at level
$(|\alpha|\beta,\alpha)$.\end{theo}
We demand that $t$ be irrational to be sure that the condition $(n',m') <
(|\alpha|\beta,\alpha)$ implies $(n',m') \in I''$ is satisfied.
\vspace{7 mm}

The values of $\jmath_0$ and $\jmath_1$ leading to a vanishing vector are
restricted by polynomial equations. Hence, we can choose $\jmath_0$ and
$\jmath_1$ almost arbitrarily to get the singular vector. We shall illustrate
this point below.

\subsection{Some matrix forms for singular vectors} \label{sec:mfsv}

In equation (\ref{eq:tri}), it is possible to put the vectors
$|n,m\rangle_{\jmath_{\alpha,\beta}(t)}$ for $(n,m) < (|\alpha|\beta,\alpha)$
together to build a column vector with
$((|\alpha|\beta+\alpha+1)(|\alpha|\beta+1)-1)$ components.
We have to choose a total ordering for the couples $(n,m) <
(|\alpha|\beta,\alpha)$.
We can even arrange things to make this total ordering compatible with the
partial ordering we had before (but there is no canonical way to do this).
We write for instance
$\vec{f}=(|\beta|\alpha|,\alpha-1\rangle_{\jmath_{\alpha,\beta}(t)},\cdots,
{|0,0\rangle}_{\jmath_{\alpha,\beta}(t)})^{tr}$ and 
$\vec{F}=(\overline{|\beta|\alpha|,\alpha\rangle}_{\jmath_{\alpha,\beta}(t)},0,
\cdots,0)^{tr}$.
Equation (\ref{eq:tri}) is then recast in a matrix form $\vec{F}={\bf M }
\vec{f}$.
The matrix elements of ${\bf M}$ are of course operators.

We shall also use the notation
$|\;\;\rangle^{(sing)}_{\jmath_{\alpha,\beta}(t)}$ for the state
$\overline{|\beta|\alpha|,\alpha\rangle}_{\jmath_{\alpha,\beta}(t)}$.
The matrix ${\bf M}$ is triangular.

\vspace{7 mm}

In certain circumstances, a simpler matrix form is available.
This is based on the truncation of the descent equations (see section
\ref{sec:Tr} and \ref{sec:fqvm}).
If $\alpha$ is positive, we choose $\jmath_0$ and $\jmath_1$ such that
$\jmath_0 -\jmath_1=\jmath_{\alpha,\beta}(t)-\alpha$ and $\jmath_0
+\jmath_1=\jmath$, i.e. $2\jmath_0=-t\beta-1$, $2\jmath_1=\alpha$.
In this case, we know that the couples $(n,m)$ with $m < 0$ or $m > \alpha$ do
not contribute.
This leads to a matrix form for the singular vector, involving only the states
$|n,m\rangle_{\jmath_{\alpha,\beta}(t)}$ with $  0 \leq m \leq \alpha$ and $0
\leq n \leq \alpha \beta$. The number of components of the vectors is reduced
to $((\alpha\beta+1)(\alpha+1)-1)$
A similar construction is also possible if $\alpha$ is negative.
To be sure that we obtain the singular vector, we ought to prove that the
values of $\jmath_0$ and $\jmath_1$ do not satisfy the fusion rules. We
conjecture that this is true.

\vspace{5 mm}

The case, when $\alpha=1$ is interesting. We remark that
$\jmath_{1,\beta}(t)=-\frac{t\beta}{2}$.
The family of equations (\ref{eq:tri}) can be restricted to
\begin{eqnarray} \label{eq:psv} tn|n,0\rangle_{-\frac{t\beta}{2}} & = &
\sum_{p=1}^nJ^+_{-p}|n-p,1\rangle_{-\frac{t\beta}{2}} +
\sum_{p=1}^nJ^0_{-p}|n-p,0\rangle_{-\frac{t\beta}{2}} \nonumber \\
t(n-\beta)|n,1\rangle_{-\frac{t\beta}{2}} & = &
-\sum_{p=1}^nJ^0_{-p}|n-p,1\rangle_{-\frac{t\beta}{2}} +
\sum_{p=0}^nJ^-_{-p}|n-p,0\rangle_{-\frac{t\beta}{2}} \end{eqnarray}
We recall that the singular vector is given by the right hand side of the
degenerate equation corresponding to the singular level $(n,m)=(\beta,1)$.
The associated matrix form can be written explicitly. We give an example in
section \ref{sec:112}.
These expressions play the same role for $A^{(1)}_1$ as do the matrix
expressions (see \cite{bfiz}) of the Benoit--Saint Aubin formul\ae\ (see
\cite{bsa}) for the Virasoro algebra. We shall comment on this in the next
section.

In this case, we have computed the overlap function (see appendix \ref{sec:of})
$\Gamma_{\beta,1}$ for $(\jmath_0,\jmath_1)=(\frac{-t\beta-1}{2},\frac{1}{2})$
and $(\jmath_0',\jmath_1')$ for small values of $\beta$. This leads to
\begin{con} \label{conj:1} When $\jmath=\jmath_{1,\beta}(t)$, a necessary
condition for fusion from $V^{(\jmath_0,t)}$ and $V^{(\jmath_0,t)}$ in
$V^{(\jmath,t)}$ to be possible is the vanishing of the polynomial
\[ \prod_{i=-\beta+2}^{\beta} (\beta(\jmath_0+\jmath_1+1) +i\jmath)
\prod_{i=-\beta}^{\beta}  (\beta(\jmath_0-\jmath_1) +i\jmath)\]
(where in these products, $i$ is restricted to have the same parity as
$\beta$).

It $t$ is irrational, the vanishing of this polynomial is also a sufficient
condition.
\end{con}

\subsection{Projection of the recursion relations}

The family of equation (\ref{eq:tri}) involves only $U({\cal E}_-)$. We have
already emphasized several times that ${\cal E}_-$, which consists of
generators of degree less than 0 with respect to the principal gradation,
contains the generators of degree less than $-1$ as an ideal.
The quotient is a commutative Lie algebra with $J^-_0$ and $J^+_{-1}$ as
generators.
Its universal enveloping algebra is still graded by $n$ and $m$, and there is a
single generator at level $(n,m)$, $(J^+_{-1})^n (J^-_0)^{n+m}$.
We can write equation (\ref{eq:tri}) in the quotient, replacing
$|n,m\rangle_{\jmath}$ by $C_{n,m}(J^+_{-1})^n (J^-_0)^{n+m}$. The coefficients
$C_{n,m}$ are complex numbers satisfying
\begin{eqnarray*}
\lefteqn{(tn+m(2\jmath+1-m))C_{n,m}= } \\ & &(-\jmath+\jmath_0+\jmath_1+m+1)
C_{n-1,m+1}  -(-\jmath+\jmath_0-\jmath_1+m-1) C_{n,m-1}
\end{eqnarray*}
The initial condition for a proper solution is $C_{0,0}=1$.
It follows from the previous considerations that, as a function of $\jmath$ for
fixed $t$, $C_{n,m}$ is rational, with poles only at the zeroes of the
contragredient form.
The residues at the poles give the fusion rules (this is a consequence of the
normalization property of the singular vectors).
The non-appearance of the spurious poles is highly non-obvious. Hence, this
innocent-looking recursion relation contains a lot of information, and it would
be of great value to be able to study it independently. We have not been able
to do so, and leave it as an open problem. This is an appropriate point to
close this section.

\section{Some comments on Hamiltonian reduction}

We make some comments related to our initial motivations.

There is a close connection between the structure of the representations of the
$A^{(1)}_1$ algebra and the Virasoro algebra. It uses quantum Hamiltonian
reduction (see for instance \cite{bo} for the quantum case and \cite{ds} for
the classical one). We recall the basic steps of the construction. The idea is
to introduce on Verma modules for $A^{(1)}_1$ a modified Virasoro algebra. From
now on, we denote by $L_m^{(S)}$ the Virasoro generators obtained by the
Sugawara construction. We set $L^{(N)}_m=L_m^{(S)}-(m+1)J^0_m$.
We observe that there is no modification for $m=-1$.
It is easy to check that
\[[L^{(N)}_m,L^{(N)}_n]=(m-n)L^{(N)}_{m+n}
+\frac{m^3-m}{12}(15-6t-6t^{-1})\delta_{m+n}\]
With respect to this new Virasoro algebra, we obtain
\[[L^{(N)}_m,J^+_{n-1}]=-(m+n)J^+_{n+m-1}\]
\[[L^{(N)}_m,J^-_{n+1}]=(m-n)J^-_{n+m+1}\]
Hence, according to (\ref{eq:qTf}) and (\ref{eq:qTfc}), $J^+(z)
=\sum_{-\infty}^{+ \infty}J^+_n  z^{-n-1}$ and $J^-(z)=\sum_{-\infty}^{+
\infty}J^-_n  z^{-n-1}$ are primary fields of respective weights $0$ and $2$.
However,
\[[L^{(N)}_m,J^0_{n}]=-nJ^0_{n+m}-\frac{t-2}{2}m(m+1)\delta_{n+m}\]
leading to
\[ [L^{(N)}_m,
J^0(z)]=((m+1)z^m+z^{m+1}\partial_z)J^0(z)-\frac{t-2}{2}m(m+1)z^{m-1}\]
Hence $J^0(z)$ is a scaling field of weight $1$ but not a primary field.

\vspace{7 mm}

If we replace the above commutators by Poisson brackets, the system becomes
classical.
If we take $J^+(z)$ as a dynamical variable, the fact that it has conformal
weight $0$ makes it possible to reduce the phase space by the constraint
$J^+(z)=1$ without loosing conformal invariance. The correct way to treat this
problem in quantum field theory is to introduce ghosts.

To the $bc$ system with commutation relations
\[\{c_m,b_n\}=\delta_{m+n} \qquad \{c_m,c_n\}=\{b_m,b_n\}=0\]
we associate a graded Fock space. There are two states at level $0$,
$|\uparrow\rangle$ and $|\downarrow\rangle$ such that
$c_0|\downarrow\rangle=|\uparrow\rangle$ and
$b_0|\uparrow\rangle=|\downarrow\rangle$. The states $|\uparrow\rangle$ and
$|\downarrow\rangle$ are annihilated by $b_n$ and $c_n$ for positive $n$. By
definition, the Fock space is the representation obtained by acting on the
states at level $0$ with any combination of the generators. The Fock space can
be turned into a representation of the Virasoro algebra by choosing an
arbitrary parameter $s$ and taking
\[L^{(G)}_m=\sum_{n=-\infty}^{+\infty} (m(s-1)-n)b_{m+n}c_{-n} \qquad {\mbox
for }\;\; m \neq 0\]
\[L^{(G)}_0=\sum_{n=1}^{+\infty} n(b_{-n}c_n+c_{-n}b_n)+\frac{s(1-s)}{2}\]
Then one can check that
\[
[L^{(G)}_m,L^{(G)}_n]=(m-n)L^{(G)}_{m+n}+\frac{m^3-m}{12}(12s(1-s)-2)
\delta_{m+n}\]
\[ [L^{(G)}_m,b_n]=(m(s-1)-n)b_{n+m} \qquad  [L^{(G)}_m,c_n]=(m(1-s
-1)-n)c_{n+m}\]
Then according to (\ref{eq:qTf}) and (\ref{eq:qTfc}), the fields
\[ b(z)=\sum_{-\infty}^{+ \infty} b_n z^{-n-s} \mbox{ and }\; \;
c(z)=\sum_{-\infty}^{+ \infty} c_n z^{-n-1+s}\]
are primary fields of weight $s$ and $1-s$ respectively.
We define the ghost number to be $1$ for $c(z)$ and $-1$ for $b(z)$.
This leads to define the ghost number operator $U$ by
$U=\sum_{n=1}^{+\infty}(c_{-n}b_n-b_{-n}c_n)-b_0c_0$. Then $[U,Q]=Q$.

\vspace{7 mm}

We can now study the tensor product of this Fock space with a highest weight
cyclic $A^{(1)}_1$-module.
The generator $J^0_0$ commutes with the Virasoro algebra (with generators
$L^{(tot)}_m=L^{(N)}_m+L^{(G)}_m$)) and can still be diagonalized in the tensor
product.
To impose a quantum analog of the constraint $J^+(z)=1$, we define
$Q=\sum_{n=-\infty}^{+\infty} c_n(J^+_{-n-1}-\delta_{n,0})$.
It is easy to check that $Q^2=0$. The operator $Q$ commutes with the Virasoro
algebra (with generators $L^{(tot)}_m=L^{(N)}_m+L^{(G)}_m$) if and only if
$s=0$.
We assume that $s=0$ in the following.
Then $Q$ is proportional to $\oint c(z)(J^+(z)-1)$ which is geometrically
well-defined, showing clearly the relation with the appropriate constraint.
Moreover, the representation of the Virasoro algebra in the tensor product has
central charge $c=13-6t-6t^{-1}$, and the eigenvalue of $L_0$ acting on
$|\jmath,t\rangle \otimes |\uparrow\rangle$ is
$h=\frac{\jmath(\jmath+1)}{t}-\jmath=\frac{(2\jmath+1-t)^2-(1-t)^2}{4t}$. This
state is clearly annihilated by the $L_n$'s for positive $n$. The fundamental
remark is that if we take $\jmath=\jmath_{\alpha,\beta}(t)$ with $\alpha$
positive, we get
\[h_{\alpha,\beta}(t)=\frac{(\alpha-t(\beta+1))^2-(1-t)^2}{4t}\]
 and these are just the weights for which the Virasoro Verma module is not
irreducible and contains a singular vector at level $\alpha(\beta+1)$.
The cohomology of $Q$ is graded by $U$, and at a given degree, the cohomology
space carries a representation of the Virasoro algebra.
Clearly, the state $|\jmath,t\rangle \otimes |\uparrow\rangle$ has ghost number
$0$ and is in the kernel of $Q$.
It is never $Q$-exact.
This is because the only states at level 0 for the Virasoro algebra are
obtained by repeated action of $J^+_{-1}$ and $b_0$ on $|\jmath,t\rangle
\otimes |\uparrow\rangle$.
But $Q$ commutes with $J^+_{-1}$ and $Qb_0|\jmath,t\rangle \otimes
|\uparrow\rangle=(J^+_{-1}-1)|\jmath,t\rangle \otimes |\uparrow\rangle$.
Hence, no finite linear combination can lead to $|\jmath,t\rangle \otimes
|\uparrow\rangle$ by application of $Q$ (we note however that the ill-defined
$-\sum_0^{\infty} (J^+_{-1})^n b_0|\jmath,t\rangle \otimes |\uparrow\rangle$
would formally do the job).
Hence the cohomology at ghost number $0$ is non-trivial.

We believe that there is no cohomology at non-zero ghost number and that if the
$A^{(1)}_1$-module is a Verma module, the cohomology at ghost number zero is a
Verma module for the Virasoro algebra.
This result probably exists already in the literature, but we have neither been
able to find it written in an accessible language for us, nor to build a proof,
although we think there should be some elementary argument.

\vspace{7 mm}

It is easy to check that a singular vector in an $A^{(1)}_1$-module tensored
with $|\uparrow\rangle$ is annihilated by $Q$. Our hope was then to prove that
the singular vectors for $A^{(1)}_1$ with $\alpha$ positive could be easily
rewritten as polynomial as in the generators of the Virasoro algebra modulo a
$Q$-exact term.
Remark that the operator $-c_0$ has a trivial cohomology and that $Q$ is the
sum of $-c_0$ and a term decreasing the eigenvalue of $J^0_0$ by one. This
ensures that a state annihilated by $Q$  which is a finite linear combination
of eigenstates of $J^0_0$ with eigenvalues greater than $\jmath$ is always
equivalent to an eigenstate of $J^0_0$ with eigenvalue $\jmath$ modulo a
$Q$-exact term. Hence the situation is not hopeless. But we have not been able
to proceed further except in very special examples.
For instance, if $\jmath=0$,
\[tL^{(N)}_{-1}|0,t\rangle=J^+_{-1}J^-_0|0,t\rangle \qquad
L^{(G)}_{-1}|\uparrow\rangle=0\]
Hence
\begin{eqnarray*} tL^{(tot)}_{-1}|0,t\rangle \otimes |\uparrow\rangle & = &
J^+_{-1}J^-_0|0,t\rangle \otimes |\uparrow\rangle \\ & = & J^-_0|0,t\rangle
\otimes |\uparrow\rangle + (J^+_{-1}-1)J^-_0|0,t\rangle \otimes
|\uparrow\rangle \\ & = & J^-_0|0,t\rangle \otimes |\uparrow\rangle
+Qb_0J^-_0|0,t\rangle \otimes |\uparrow\rangle \end{eqnarray*}
showing that this particular singular vector for $A^{(1)}_1$  flows to the
singular vector for the Virasoro algebra under Hamiltonian reduction.
If we could do this more systematically, we would probably understand much
better the construction (see \cite{bfiz}) of singular vectors in Virasoro Verma
modules. The special case $\alpha=1$ is promising and interesting because the
relation with the Benoit--Saint-Aubin formulae (see \cite{bsa}), but has
nevertheless eluded us.

\vspace{7 mm}

Moreover, a precise solution to these question would give an interesting
shortcut for the usual proof (see \cite{bo} and the for the mathematically
inclined reader \cite{ff2}) that Hamiltonian reduction relates the minimal
models for the $A^{(1)}_1$ and the Virasoro algebra.
The usual method is quite indirect and involves bosonization in two places,
with the necessity of introducing other $Q$ operators.
A direct proof would be much more illuminating.
We leave this as an open problem.

\section{Conclusions and remarks}

The interplay between fusion, fusion rules and singular vectors has be used to
construct these singular vectors explicitly.
It is not clear for us whether these expressions can be used in other
theoretical applications, but we think that the relationship between these
aspects, although not unexpected, was not recognized to be so intimate.
The proper interpretation of the Knizhnik--Zamolodchikov equation in our
context has been of great importance.
On the other hand unitarity played no role in our discussion.
Some fusion rules have been computed, and a general calculation should be
possible.
However, many questions remain open.
Among these we would like to emphasize two.

The generalization to other affine algebras would be interesting.
There are serious technical difficulties, but they should not be insuperable.
Much more intricate seems to be the extension to other chiral algebras.
The Virasoro algebra is an example which still needs to be better understood,
and we are back to Hamiltonian reduction.

We have concentrated on purely algebraic aspects, but geometry certainly plays
a fundamental role.
We have some hints that a geometrical interpretation of the formul\ae\
(\ref{eq:psv}) exists, and is related to the analogous geometrical
interpretation of the Benoit--Saint-Aubin formul\ae\ in terms of covariant
differential equations given in \cite{bfiz}, inspired by \cite{ds}.
We observe that the two cases are related by Hamiltonian reduction.

We hope that these questions will motivate further work.

\vspace{10 mm}

\noindent {\large \bf Acknowledgements}

We benefited from many fruitful discussions with Denis Bernard and Giovanni
Felder. They helped us to understand the links between our approach of fusion
and the more mathematical one. They also suggested improvements and raised many
questions, concerning for instance the fusion rules. It is a pleasure to thank
them warmly. We also thank Claude Itzykson for a careful reading of the
manuscript.

\appendix

\section{The singular vector at level $(1,1)$.}  \label{sec:11}

The singular vector for $(\alpha,\beta)=(1,1)$ and $\jmath=-\frac{t}{2}$ is the
simplest non-trivial singular vector. We compute it in two different ways.

\subsection{The method of Malikov, Feigin and Fuks} \label{sec:111}

To illustrate the technique of analytic continuation, we do the calculation in
detail for $(\alpha,\beta)=(1,1)$. So, we are trying to make sense of
\[ (J^-_0)^{1+t} J^+_{-1} (J^-_0)^{1-t}\]
The fact that $J^+_{-1}$ already appears raised to an integral power (in fact
1) makes the situation comparatively easy. However, the general computation
follows analogous patterns. The starting point is the identity
\[  e^{xJ^-_0} J^+_{-1}e^{-xJ^-_0}= J^+_{-1}-2xJ^0_{-1}-x^2J^-_{-1} \]
which is proved for instance by differentiation.
Then we expand
\[ e^{xJ^-_0} J^+_{-1}= (J^+_{-1}-2xJ^0_{-1}-x^2J^-_{-1})e^{-xJ^-_0} \]
in powers of $x$ to get
\[ (J^-_0)^p  J^+_{-1} =J^+_{-1} (J^-_0)^p-2pJ^0_{-1}
(J^-_0)^{p-1}-p(p-1)J^-_{-1} (J^-_0)^{p-2} \qquad p=0,1,\cdots\]
We observe that the coefficients are polynomial in $p$, and we extend these
identities for complex $p$. Both sides are ill-defined. We take $p=1+t$ and
multiply the identity by $(J^-_0)^{1-t}$ on the right. This leads to
\[(J^-_0)^{1+t} J^+_{-1} (J^-_0)^{1-t} =(J^+_{-1} (J^-_0)^{1+t}-2(1+t)J^0_{-1}
(J^-_0)^{t}-t(1+t)J^-_{-1} (J^-_0)^{t-1})(J^-_0)^{1-t}\]
If we assume that the usual rules for multiplication of powers of $J^-_0$ can
be extended to complex powers, we end up with
\[(J^-_0)^{1+t} J^+_{-1} (J^-_0)^{1-t}=J^+_{-1}(J^-_0)^2-2(1+t))J^0_{-1} J^-_0
-t(1+t)J^-_{-1}\]
The right hand side gives a definition of the left hand side. We remark that
the left hand side was already well-defined for $t\in \{-1,0,1\}$. It is easy
to check that at these special values, the two definitions coincide.

Of course, we could have started with an identity for $ J^+_{-1}(J^-_0)^p$. We
do not prove that the result is the same. This is a consequence of the general
theory of Malikov, Feigin and Fuks \cite{mff}.

It is clear that even when $\alpha=1$, if $\beta > 1$ formula (\ref{eq:mff1})
contains more factors, making the computation more and more complicated. This
is to be contrasted with the form given in equation (\ref{eq:psv}).

\subsection{The matrix form} \label{sec:112}

In the case when $(\alpha,\beta)=(1,1)$, our method leads to the following
computation. The family of equations (\ref{eq:psv}) reduces to
\begin{eqnarray*}-t|0,1\rangle_{-\frac{t}{2}} & = &
J^-_{0}|0,0\rangle_{-\frac{t}{2}}\\
 t|1,0\rangle_{-\frac{t}{2}} & = & J^+_{-1}|0,1\rangle_{-\frac{t}{2}} +
J^0_{-1}|0,0\rangle_{-\frac{t}{2}}\\
|\;\;\rangle^{(sing)}_{-\frac{t}{2}} & = & -J^0_{-1}|0,1\rangle_{-\frac{t}{2}}
+ J^-_{0}|1,0\rangle_{-\frac{t}{2}}+J^-_{-1}|0,0\rangle_{-\frac{t}{2}}
\end{eqnarray*}
It is easy to recast this in a matrix form. We write
\[ \left(\begin{array}{c} |\;\;\rangle^{(sing)}_{-\frac{t}{2}} \\ 0 \\ 0
\end{array}\right) =\left(\begin{array}{ccc} J^-_0 & -J^0_{-1} & J^-_{-1} \\ t
& -  J^+_{-1} & -J^0_{-1} \\ 0 & t & J^-_{0} \end{array}\right)
\left(\begin{array}{c}  |1,0\rangle_{-\frac{t}{2}}  \\
|0,1\rangle_{-\frac{t}{2}}\\ |0,0\rangle_{-\frac{t}{2}}\end{array}\right)\]
We solve this triangular system and obtain
\[ |\;\;\rangle^{(sing)}_{-\frac{t}{2}}=-\frac{1}{t^2} \left(J^-_{0} J^+_{-1}
J^-_{0}-t( J^-_{0} J^0_{-1}+ J^-_{0} J^0_{-1}) + J^-_{-1}\right)
|0,0\rangle_{-\frac{t}{2}}\]
Using the commutation relations to rewrite the right hand side of this equation
in the basis (\ref{eq:pbw}), it is easy to check that the different expressions
for the singular vector are proportional to each other.
The analogous computations for $\beta > 1$ become more and more tedious, but
they are much simpler that the ones involved in the computation by analytic
continuation.
There is some intuitive explanation for this: our recursion formul\ae\ define
the singular vector, without specifying a basis of $U({\cal E}_-)$, with the
consequence that in a sense ``the singular vector itself chooses the way it
wants to be expressed''.

\section{Further covariance constraints} \label{sec:pi}

We are going to study the covariance properties of the state (\ref{eq:pi}) of
section \ref{sec:fde} with respect to the current algebra. So we apply our
method to the state
\begin{equation}  e^{zL_{-1}+xJ^-_0}\Phi_{\jmath_0}(-z,-x)|\jmath_1,t\rangle
\end{equation}
The left ideal annihilating  $|\jmath_1,t\rangle$ is generated by
$J^0_0-\jmath_1$,
$k-(t-2)$,
$L_0-h_1$,
and the $J^a_n$'s in ${\cal E}_+$.
We use once more the commutators (\ref{eq:comm}) and then conjugate with
$e^{zL_{-1}+xJ^-_0}$ to get
\[ e^{zL_{-1}+xJ^-_0}(J^0_0-D^0_{\jmath_0}-\jmath_1)  e^{-zL_{-1}-xJ^-_0} \;\;
e^{zL_{-1}+xJ^-_0}\Phi_{\jmath_0}(-z,-x)|\jmath_1,t\rangle=0 \]
\[ e^{zL_{-1}+xJ^-_0}(L_0 -h_1 -z\partial_z-h_0) e^{-zL_{-1}-xJ^-_0} \;\;
e^{zL_{-1}+xJ^-_0}\Phi_{\jmath_0}(-z,-x)|\jmath_1,t\rangle =0\]
\[ e^{zL_{-1}+xJ^-_0}(J^a_n-(-)^{n+a}z^n D^a_{\jmath_0}) e^{-zL_{-1}-xJ^-_0}
\;\; e^{zL_{-1}+xJ^-_0}\Phi_{\jmath_0}(-z,-x)|\jmath_1,t\rangle =0   \;\;
\forall J^a_p \in {\cal E}_+ \]

\vspace{7 mm}

We can now prove
\begin{lem} \label{lem:icc} The covariance constraints on (\ref{eq:p}) and
(\ref{eq:pi}) coincide. \end{lem}
We use the commutation relations between the stress-energy tensor and the
currents to check that
\[ e^{zL_{-1}+xJ^-_0}(J^0_0-D^0_{\jmath_0}-\jmath_1)
e^{-zL_{-1}-xJ^-_0}=J^0_0-D^0_{\jmath_1}-\jmath_0\]
\[ e^{zL_{-1}+xJ^-_0}(L_0 -h_0 -z\partial_z-h_1) e^{-zL_{-1}-xJ^-_0} =L_0 -h_0
-z\partial_z-h_1\]
It is quite tedious to show directly that the operators
\[ e^{zL_{-1}+xJ^-_0}(J^a_n-(-)^{n+a}z^n D^a_{\jmath_0}) e^{-zL_{-1}-xJ^-_0} \]
for $J^a_n$ in ${\cal E}_+$ are ($(z,x)$ dependent) linear combinations of the
operators appearing in the constraints for
$\Phi_{\jmath_1}(z,x)|\jmath_0,t\rangle$.
Happily,
as we emphasized above,
two particular constraints generate them all.
So we are left with two simple computations
\[ e^{zL_{-1}+xJ^-_0}(J^-_{-1}-z D^-_{\jmath_0}) e^{-zL_{-1}-xJ^-_0}
=J^-_{-1}-z D^-_{\jmath_1}\]
and
\[ e^{zL_{-1}+xJ^-_0}(J^+_0+D^+_{\jmath_0}) e^{-zL_{-1}-xJ^-_0}
=J^+_0-2xJ^0_0+2\jmath_0 x -x^2\partial_x=
(J^+_0-D^+_{\jmath_1})-2x(J^0_0-D^0_{\jmath_1}-\jmath_0)\]
This concludes the proof that the covariance constraints on (\ref{eq:p}) and
(\ref{eq:pi}) are the same.$\Box$

It is in this sense that we can identify these two states.

\section{Fusion of quotients of Verma modules} \label{sec:fqvm}

We give an interpretation of lemma \ref{lem:tru}.
This will also  lead to some illustrations of the comments we made after the
definition of fusion.
This section is very close in spirit to the computation of the fusion rules in
\cite{zf}.
We note that if $\jmath_1$ is a nonnegative integer or half-integer,
$(J^-_0)^{2\jmath_1+1}|\jmath_1,t\rangle$ is a singular vector in
$V^{(\jmath_1,t)}$.
This singular vector generates a submodule,
and we can take the quotient.
In this quotient the left ideal of $U(\hat{\cal A})$ annihilating
$|\jmath_1,t\rangle$ contains $(J^-_0)^{2\jmath_1+1}$.
So if we try to implement fusion of this quotient module with
$V^{(\jmath_0,t)}$ to get $V^{(\jmath,t)}$ there is a new constraint.
\begin{lem} This constraint is simply that
\[\partial_x^{2\jmath_1+1} \Phi_{\jmath_1}(z,x)|\jmath_0,t\rangle =0 \]
where this time $\Phi_{\jmath_1}$ stands for the primary field associated to
the quotient module.\end{lem}
We multiply the relation $(J^-_0)^{2\jmath_1+1}|\jmath_1,t\rangle=0$ on the
left by $ e^{zL_{-1}+xJ^-_0}\Phi_{\jmath_0}(-z,-x)$
Then a simple application of the commutation relations (\ref{eq:comm}) leads to
\[\partial_x^{2\jmath_1+1}
e^{zL_{-1}+xJ^-_0}\Phi_{\jmath_0}(-z,-x)|\jmath_1,t\rangle =0 \]
But, as far as covariance is concerned,
we have shown that it is possible to identify
$e^{zL_{-1}+xJ^-_0}\Phi_{\jmath_0}(-z,-x)|\jmath_1,t\rangle$ with
$\Phi_{\jmath_1}(z,x)|\jmath_0,t\rangle$.$\Box$

\vspace{5 mm}

{}From this we deduce that in the $\jmath$-sector
\[\partial_x^{2\jmath_1+1}  \sum_{n,m} z^{h-h_0-h_1+n}
x^{\jmath_0+\jmath_1-\jmath+m} |n,m\rangle_{\jmath}=0\]
Hence for any $(n,m)\in I$
\begin{equation} \label{eq:fr1} (\jmath_0+\jmath_1-\jmath+m)
(\jmath_0+\jmath_1-\jmath+m-1) \cdots
(\jmath_0+\jmath_1-\jmath+m-2\jmath_1)|n,m\rangle_{\jmath}=0 \end{equation}
In particular, either $|0,0\rangle_{\jmath}=0$ or $\jmath \in
\{\jmath_0+\jmath_1,\jmath_0+\jmath_1-1,\cdots,\jmath_0-\jmath_1\}$.
This is a fusion rule. It looks quite familiar. If we define $i_+$ and $i_-$ by
$\jmath_1=1/2(i_+-i_-)$, $\jmath_0-\jmath=1/2(i_++i_-)$ then the content of
(\ref{eq:fr1}) is equivalent to the truncations of the descent equations
obtained in \ref{lem:tru}.
We observe that the use of the quotient module of $V^{(\jmath_1,t)}$ to define
fusion imposes that the operator product expansion has no singularity in
$x$-space.

\vspace{7 mm}

What if $\jmath_0$ is a nonnegative integer or half-integer?
We can guess that the consequence of the  existence of the singular vector in
$V^{(\jmath_0,t)}$ is a fusion rule $\jmath \in
\{\jmath_1+\jmath_0,\jmath_1+\jmath_0-1,\cdots,\jmath_1-\jmath_0\}$.
This is indeed the case.
\begin{lem} In the fusion of $V^{(\jmath_1,t)}$ with the quotient module of
$V^{(\jmath_0,t)}$, the new constraint is
\[(J^-_0-\partial_x)^{2\jmath_0+1}\Phi_{\jmath_1}(z,x)|\jmath_0,t\rangle =0\]
\end{lem}
Starting from $(J^-_0)^{2\jmath_0+1}|\jmath_0,t\rangle =0$, which holds in the
quotient module of $V^{(\jmath_0,t)}$,
this is a simple application of the commutation relations
(\ref{eq:comm}).$\Box$

\vspace{5 mm}

We deduce that in the $\jmath$-sector
\[(J^-_0-\partial_x)^{2\jmath_0+1} \sum_{n,m} z^{h-h_0-h_1+n}
x^{\jmath_0+\jmath_1-\jmath+m} |n,m\rangle_{\jmath}=0\]
There is only one term belonging to $V_{0,0}$, namely
\[(-\partial_x)^{2\jmath_0+1} z^{h-h_0-h_1}x^{\jmath_0+\jmath_1-\jmath}
|0,0\rangle_{\jmath}\]
{}From this we infer
\[ (\jmath_1+\jmath_0-\jmath) (\jmath_1+\jmath_0-\jmath-1) \cdots
(\jmath_1+\jmath_0-\jmath-2\jmath_0)|0,0\rangle_{\jmath}=0\]
This leads to the expected fusion rule.

\vspace{7 mm}

We turn now to the case when $t/2-\jmath_0-1$ is a nonnegative integer or
half-integer. It is easy to check that
$(J^+_{-1})^{t-2\jmath_0-1}|\jmath_0,t\rangle$ is a singular vector in
$V^{(\jmath_0,t)}$. If we use the quotient module for fusion, we get a new
constraint.
\begin{lem}  In the fusion of $V^{(\jmath_1,t)}$ with the quotient module of
$V^{(\jmath_0,t)}$, the new constraint is
\[ (J^+_{-1}-z^{-1}D^+_{\jmath_1})^{t-2\jmath_0-1}
\Phi_{\jmath_1}(z,x)|\jmath_0,t\rangle =0 \]\end{lem}
 The proof is similar to the above ones.$\Box$

\vspace{5 mm}

We deduce that in the $\jmath$-sector
\[(J^+_{-1}-z^{-1}D^+_{\jmath_1})^{t-2\jmath_0-1} \sum_{n,m} z^{h-h_0-h_1+n}
x^{\jmath_0+\jmath_1-\jmath+m} |n,m\rangle_{\jmath}=0\]
There is only one term belonging to $V_{0,0}$, namely
\[
(-z^{-1}D^+_{\jmath_1})^{t-2\jmath_0-1}z^{h-h_0-h_1}
x^{\jmath_0+\jmath_1-\jmath} |0,0\rangle_{\jmath}\]
{}From this we infer
\[(\jmath_1-\jmath_0+\jmath)(\jmath_1-\jmath_0+\jmath-1) \cdots
(\jmath_1-\jmath_0+\jmath-t+2\jmath_0+2)|0,0\rangle_{\jmath}=0\]
and this leads to a fusion rule.
The spin $\jmath$ has to belong to
$\{t-2-\jmath_0-\jmath_1,t-2-\jmath_0-\jmath_1-1,\cdots,\jmath_0-\jmath_1\}$.
This is of course also familiar. We can guess the analogous fusion rule when
$t/2-\jmath_1-1$ is a nonnegative integer or half-integer.
In this case, the spin $\jmath$ has to belong to
$\{t-2-\jmath_1-\jmath_0,t-2-\jmath_1-\jmath_0-1,\cdots,\jmath_1-\jmath_0\}$.

\vspace{7 mm}

Putting all these results together, we obtain the usual conditions for fusion.
\begin{itemize}
\item If both  $\jmath_0$ and $\jmath_1$ are nonnegative integers or
half-integers, we simply recover the law of composition of spins.
The spin $\jmath$ has to belong to
$\{\jmath_1+\jmath_0,\jmath_1+\jmath_0-1,\cdots,|\jmath_1-\jmath_0|\}$.
Thus it too is an integer or half-integer, and $V^{(\jmath,t)}$ contains a
singular vector. We have only obtained a necessary condition for fusion to be
possible. But it is not difficult to show that the fusion involving the three
quotient is possible and unique if $t$ is irrational. We do not give the proof
here.
\item If moreover $t-2$ is a positive integer, we recover the full set of
fusion rules for the unitary models. Note that unitarity played no direct role
in the discussion. This is common in the representation theory of finite
dimensional semi-simple Lie algebras, where the requirement of finite
dimensionality of a representation implies its unitarity.
\end{itemize}

Let us stress once more that, although our definitions did not prevent short
distance singularities in $x$-space, these singularities disappear when we
consider fusion of quotients of appropriate Verma modules.

\section{The overlap function} \label{sec:of}

In the definition of fusion, $\jmath_0$ and $\jmath_1$ play the role of
parameters,  and it is interesting to have some kind of measure of how much the
solutions of the descent equations differ at level $(n,m)$ when $\jmath_0$ and
$\jmath_1$ vary. The contragredient form gives such a ``measure''.
We define the ``overlap'' between two solution of the descent equations,
corresponding to distinct couples $(\jmath_0,\jmath_1)$, and
$(\jmath_0',\jmath_1')$ but of course with the same value of $\jmath$ and $t$
to be
\[\Gamma_{n,m}(\jmath_0,\jmath_1,\jmath_0',\jmath_1',\jmath,t) \equiv
_{\jmath}\langle n,m| n,m\rangle'_{\jmath}\]

Using the method of sections \ref{sec:frde} and \ref{sec:srde}, it is easy to
show that the overlap satisfies recursion relations.

\begin{lem} \label{eq:ro} The overlap $\Gamma_{n,m}$ satisfies
\begin{eqnarray} (tn+m(2\jmath+1-m))\Gamma_{n,m} & = &
(-\jmath+\jmath_0+\jmath_1+m+1) (-\jmath+\jmath_0'+\jmath_1'+m+1)  \sum_{p=1}^n
 \Gamma_{n-p,m+1} \nonumber \\ & + &
2(-\jmath+\jmath_0+m)(-\jmath+\jmath_0'+m)\sum_{p=1}^n \Gamma_{n-p,m}\\ & + &
 (-\jmath+\jmath_0-\jmath_1+m-1) (-\jmath+\jmath_0'-\jmath_1'+m-1)\sum_{p=0}^n
\Gamma_{n-p,m-1} \nonumber \end{eqnarray} \end{lem}
To prove this, we first use (\ref{eq:tri}) for $|n,m\rangle_{\jmath}'$, and
then we use the descent equations for $_{\jmath} \langle n,m|$. This procedure
in not symmetric, but the final formula treats $(\jmath_0,\jmath_1)$ and
$(\jmath_0',\jmath_1')$ in a symmetric way.$\Box$

These relations are quite complicated as they stand, but by using truncation,
it is possible to use them to compute for instance fusion rules.

Theorem \ref{theo:fin} allows us to say something about the structure of the
overlap $\Gamma_{|\alpha|\beta,\alpha}$ when $\jmath=\jmath_{\alpha,\beta}(t)$.
In fact, for these very special values of the indices, the right hand side of
(\ref{eq:ro}) has to split as a product of the fusion rules for
$(\jmath_0,\jmath_1)$ and $(\jmath_0',\jmath_1')$. This is because
$\overline{|\beta|\alpha|,\alpha\rangle}_{\jmath_{\alpha,\beta}(t)}$ and
$\overline{|\beta|\alpha|,\alpha\rangle'}_{\jmath_{\alpha,\beta}(t)}$ are both
proportional to the singular vector, and the right hand side of (\ref{eq:ro})
computes the obstruction to the solving of the descent equation at level
$(|\alpha|\beta,\alpha)$.

Hence, the overlap equation provides a method to compute the fusion rules. When
$\alpha=1$ and $\jmath=\jmath_{1,\beta}(t)=-\beta \frac{t}{2}$ for instance, it
is possible to compute $\Gamma_{\beta,1}$ for small values of $\beta$, taking
$(\jmath_0,\jmath_1)=(\frac{-t\beta-1}{2},\frac{1}{2})$ (these are the values
leading to the truncation of the descent equations) and
$(\jmath_0',\jmath_1')$. This leads to conjecture \ref{conj:1}.

\end{document}